\documentclass[journal]{IEEEtran}
\usepackage[cmex10]{amsmath}
\usepackage[cmex10]{amsmath}

\usepackage{ifpdf}

\usepackage{algorithm}
\usepackage{algpseudocode}
\usepackage{xcolor}
\usepackage{url}
\ifpdf
\usepackage[pdftex]{graphicx}
\else
\usepackage[dvips]{graphicx}
\fi
\usepackage{tabularx}
\usepackage{multirow}
\graphicspath{{graphics/png/}}
\DeclareGraphicsExtensions{.png, .eps, .pdf}
\usepackage{float} 
\usepackage{subfigure}
\usepackage{comment}
\usepackage{bm}
\usepackage{amsmath,amsthm,amsfonts,amssymb}
\usepackage{pifont}
\usepackage[noadjust]{cite}
\usepackage{nomencl}
\usepackage{hyperref}

\theoremstyle{definition}
\newtheorem{theorem}{\textbf{Theorem}}

\newtheorem{remark}[theorem]{\textbf{Remark}}

\newcommand{\up}[1]{^\mathrm{#1}}
\newcommand\Tstrut{\rule{0pt}{2.6ex}}         % = `top' strut
\newcommand{\rev}[1]{\textcolor{black}{#1}}

\IEEEoverridecommandlockouts
\begin{document}
%
% paper title
% Titles are generally capitalized except for words such as a, an, and, as,
% at, but, by, for, in, nor, of, on, or, the, to and up, which are usually
% not capitalized unless they are the first or last word of the title.
% Linebreaks \\ can be used within to get better formatting as desired.
% Do not put math or special symbols in the title.
\title{Energy Storage State-of-Charge Market Model}

%% To specify the authors when (number of affiliations <= 2)
\author{
    Ningkun Zheng,~\IEEEmembership{Student Member,~IEEE,} 
    Xin Qin,~\IEEEmembership{Student Member,~IEEE,} 
    Di Wu,~\IEEEmembership{Senior Member,~IEEE,}\\
    Gabe Murtaugh,
    Bolun~Xu,~\IEEEmembership{Member,~IEEE}
	\thanks{This work was supported in part by Columbia University Data Science Institute Seed Grant UR01006728, and in part by the U.S. Department of Energy, Office of Electricity through the Energy Storage program under Contract DE-AC05-76RL01830.
    
    N.~Zheng and B.~Xu are with Columbia University, NY, USA (e-mail: nz2343@columbia.edu, bx2177@columbia.edu); X.~Qin is with the University of Cambridge, UK (e-mail: xq234@cam.ac.uk); D.~Wu is with Pacific Northwest National Laboratory, WA, USA (e-mail: di.wu@pnnl.gov); G.~Murtaugh is with California ISO, CA, USA (e-mail: gmurtaugh@caiso.com ). 

}
}

%% To specify the authors when (number of affiliations > 2)
% \author{\IEEEauthorblockN{Author n.1\IEEEauthorrefmark{1},
% Author n.2\IEEEauthorrefmark{2},
% Author n.3\IEEEauthorrefmark{3}, 
% Author n.4\IEEEauthorrefmark{3} and
% Author n.5\IEEEauthorrefmark{4}}
% \IEEEauthorblockA{\IEEEauthorrefmark{1} Department Name of Organization A\\
% Name of the organization A,
% Address A\\ Emails if wanted}
% \IEEEauthorblockA{\IEEEauthorrefmark{2} Department Name of Organization B\\
% Name of the organization B,
% Address B\\ Emails if wanted}
% \IEEEauthorblockA{\IEEEauthorrefmark{3} Department Name of Organization C\\
% Name of the organization C,
% Address C\\ Emails if wanted}
% \IEEEauthorblockA{\IEEEauthorrefmark{4}Department Name of Organization D\\
% Name of the organization D,
% Address D\\ Emails if wanted}
% }

% make the title area
\maketitle

% As a general rule, do not put math, special symbols or citations
% in the abstract
\begin{abstract}
This paper introduces and rationalizes a new model for bidding and clearing energy storage resources in wholesale energy markets. Charge and discharge bids in this model depend on the storage state-of-charge (SoC). In this setting, storage participants submit different bids for each SoC segment. The system operator monitors the storage SoC and updates their bids accordingly in market clearings. Combined with an optimal bidding design algorithm using dynamic programming, our paper shows that the SoC segment market model provides more accurate representations of the opportunity costs of energy storage compared to existing power-based bidding models. The new model also captures the inherent SoC-dependent operational characteristics of energy storage. We benchmark the SoC segment market model against an existing single-segment model in price-taker and price-influencer simulations. \rev{The simulation results show that compared to the existing power-based bidding model, the proposed model improves profits by 10-56\% in the price-taker case study; the model also improves total system cost reduction from storage by around 5\%, and helps reduce price volatilities in the price-influencer case study.}

% Our price-taker simulation results show that the proposed model provides a 10\% to 60\% improvement in profitability compared to the existing storage model. Our price-influencer simulation results show that the proposed model provides around 5\% more system cost reduction compared to the existing storage model, and is significantly better at reducing price volatilities.
\end{abstract}


\IEEEpeerreviewmaketitle
% Use this to place sponsorships
% \thanksto{Applicable sponsors, if any, should be placed using the \emph{thanksto} command.}

\section{Introduction}
Energy storage resources, especially battery energy storage, are entering wholesale electricity markets at a surging rate. \rev{The battery capacity connected to the California Independent System Operator (CAISO), the power system operator and market organizer of the state of California, has increased from 488~MW at the end of 2020 to 4,367~MW as of Sep 2022 and is expected to reach near 10~GW in 2026~\cite{caiso-1,caiso-2,caiso-3}. According to EIA Annual Electric Generator Report, with increasingly installed energy storage capacity flatten the ancillary service market price, majority of energy storage participants starting to focus on arbitraging in wholesale energy markets~\cite{us2021form}}.

Integrating energy storage resources into wholesale electricity markets requires the development of new models. In centralized electricity markets, which cover most regions in North America, participants must bid using a resource model representing their operational characteristics in the market clearing. For example, thermal generators submit power segment bids stemming from their heat rate curves and other operation or cost parameters, including startup costs, no-load costs, ramp rates, and minimum up and down time~\cite{kirschen2018fundamentals}. However, energy storage resources have distinctly different operational characteristics compared to thermal generators and need different bidding parameters. The FERC (Federal Energy Regulatory Commission) has recognized this need and issued Order 841, which requires that future electricity market designs ``account for the physical and operational characteristics of electric storage resources through bidding parameters or other means''~\cite{federal2018electric}. 

Managing storage state-of-charge (SoC) is critical for energy storage participants. \rev{The storage opportunity cost depends on SoC, and various storage operation factors, including degradation rates and efficiencies, depend on power rating and SoC~\cite{xu2020operational, xu2017factoring, pandvzic2018accurate}.} Managing SoC is achievable in day-ahead markets with a 24-hour optimization horizon but is not effective in real-time markets~\cite{fang2022efficient}. Currently, most system operators allow storage to participate in markets by self-scheduling or submitting both charge and discharge bids~\cite{sakti2018review}. Storage has complete control over its SoC in self-scheduling but loses the flexibility to react to actual system conditions, decreasing revenue potential and social welfare.  Managing SoC through existing bidding models is difficult as storage participants cannot update their bids in a timely manner based on SoC~\cite{bhattacharjee2022energy}.

This paper introduces a storage market model that allows storage participants to effectively manage their economic bids based on SoC and better represents SoC-dependencies of energy storage physical and operational characteristics.
% state-of-charge segment single-period economic dispatch model for energy storage participation in the real-time electricity markets. The model allows energy storage to submit charge and discharge bids by segments according to the SoC ranges. This model also allows storage to specify different power rating and efficiency parameters for each SoC segment. 
% This model addresses the integrating opportunity cost issues when energy storage under an economic dispatch model without needing a look-ahead window. Our method can further model energy storage with nonlinear physical parameters with a mixed-integer linear programming (MILP) model by introducing integer variables to SoC segments while providing high computational efficiency and close to maximum potential profit.
The main contribution is five-fold:
\begin{itemize}
    \item We introduce an SoC segment market model for energy storage participation to \emph{economically} manage their SoC in wholesale electricity markets. The model allows energy storage to submit power rating, efficiency, and charge and discharge bids by segments according to the SoC ranges. % This model also allows storage to specify different power rating and efficiency parameters for each SoC segment.
    \item \rev{We incorporate SoC-dependent physical parameters into our previous storage bidding algorithm to generate time-varying SoC-dependent charge and discharge bid curves.} 
%     \item In an effort to model nonlinear physical parameters in energy storage, we extend the linear dispatch model to a MILP model by introducing integer variables to constraint SoC segment charge/discharge order.
    \item \rev{We combine the proposed market model with the optimal bidding algorithm to benchmark the proposed model with existing market models in terms of system costs, price results, and storage revenue in price-taker and price-influencer cases.}
    \item For storage models whose parameters are \emph{independent} of SoC, we model SoC-dependent bids as linear programming in real-time markets. Results show the proposed model reduces total system costs and price volatility with little effect on the computation time.
    \item For storage models whose physical parameters are \emph{dependent} of SoC, we model SoC-dependent bids using mixed-integer linear programming, in which integer variables are required to model the inherent non-convexity. Results show modeling the SoC dependency in day-ahead (multi-period) and real-time markets can significantly reduce system costs and price volatilities, and improve storage revenue, but have to trade-off computation time.
\end{itemize}

We organize the remainder of the paper as follows. Section~\ref{LR} reviews related literature. Sections~\ref{form} and~\ref{bidmodel} present the proposed market model and the bidding algorithm. Sections~\ref{ptcs} and~\ref{pics} describe case studies under price-taker and price-influencer settings. Section~\ref{conc} concludes this paper.

\section{Literature Review}\label{LR}

\subsection{Energy Storage Market Models}\label{sec:2a}

Independent system operators and regional transmission organizations (ISOs/RTOs) across North America are implementing new market rules to reduce barriers to energy storage participation, facilitated by FERC Order 841~\cite{konidena2019ferc}. In current and upcoming market designs, most system operators are allowing storage to bid as a combination of a generator and a flexible load in day-ahead or real-time markets~\cite{reviewall,sakti2018review}. Dispatching energy storage using market bids requires marginal changes to the market clearing model as it simply combines two existing market models. The upcoming market design can schedule storage in day-ahead markets since the unit commitment model used to clear the market considers a 24-hour operation horizon, through which the storage SoC can be optimally managed using temporal constraints~\cite{caiso_es}. 

Distinct from the day-ahead market, generating resources in real-time energy markets are cleared 5 minutes ahead of the dispatch. Real-time market prices are more volatile than day-ahead due to more constrained generator status and demand forecast errors. Therefore, real-time markets may offer higher arbitrage profits for storage~\cite{xu2017factoring,byrne2020opportunities}, and real-time dispatch better utilizes storage's near-instantaneous response speed to compensate for demand and renewable fluctuations. However, incorporating storage in real-time markets is non-trivial. In most North American systems, real-time markets only consider a single time step, while CAISO and the New York Independent System Operator include a short look-ahead horizon to address commitment and ramping constraints. This look-ahead horizon is too short to capture daily demand and price patterns to effectively manage storage SoC constraints while incorporating a longer horizon increases the computation cost significantly and may fail to meet the 5-minute dispatch frequency~\cite{zhao2019multi,caiso_es}.

A key challenge of incorporating storage into real-time markets is quantifying the opportunity cost. The system operator cannot incorporate a sufficiently long look-ahead horizon into real-time dispatch. For example, a battery owner will have to determine its charge bid prices based on predictions of the price at which the charged energy will be sold later.
% Different types of profit-maximizing arbitrage strategies have been proposed from the battery owner's perspective. From the perspective of market designs, Chen et al.~\cite{9chen2021pricing} proposes a novel pricing mechanism to incorporate storage opportunity costs in electricity prices, which encourages storage owners to truthfully bid their marginal costs. 
Many studies have explored how different participation strategies could impact storage revenue and market efficiencies, including self-scheduling~\cite{12akbari2019hybrid, mohsenian2015coordinated} and bidding~\cite{10shuai2018optimal, 13jiang2015optimal, 4krishnamurthy2017energy}. Some of these studies have concluded that considering SoC in bid designs is critical to the storage market revenue~\cite{bhattacharjee2022energy, 7wang2017look, gao2022multiscale}. However, few studies have investigated alternative market designs to better manage SoC through bid parameters, which is the problem targeted in this paper.

% From the perspective of storage owners, different types of arbitrage strategies have been proposed to maximize storage profits in real-time markets. 
% The first type is self-scheduling, in which a storage owner only submits discharging and charging power without prices to ISOs. For example, Akbari et al.~\cite{12akbari2019hybrid} developed a robust optimization method for storage owners under real-time price uncertainty to guarantee a certain amount of profits at the expense of losing part of arbitrage opportunities. Shuai et al.~\cite{10shuai2018optimal} incorporated battery models into a microgrid bidding strategy in real-time markets. The second type is quantity-price bids, in which a storage owner submits an offer curve to the market, similar to how thermal generators bid. Xu et al.~\cite{xu2020operational} and Jiang et al.~\cite{13jiang2015optimal} proposed novel variants of dynamic programming methods to efficiently incorporate storage opportunity costs into real-time bidding.  The approaches and effects of incorporating quantity-price bids with the wholesale electricity market models is remaining to be explored.

% all have neglected the market power of storage over market prices, and therefore, only apply to scenarios in which a power system has very few storage participants. 

\subsection{\rev{Storage Parameters State-of-Charge Dependency}}\label{socdep}

\rev{In practice, energy storage parameters, including power rating, efficiency, and discharge cost, often have nonlinear relationships with storage SoC  for various reasons based on the technology, such as the voltage dependency in electrochemical batteries~\cite{ecker2014calendar} and storage pressure levels in compressed air energy storage~\cite{calero2019compressed}. Utility-scale energy storage systems in the US are primarily Li-ion batteries with a 4-hour duration (.25 C-rate). According to lab test data, operation power rating has a limited impact on energy storage parameters at a low C-rate~\cite{saxena2019accelerated,jafari2019improved}, and SoC has the highest influence in utility-scale Li-ion battery degradation~\cite{kim2022comparison}. Therefore, the dependencies of power rating are not in the scope of this paper, but we can incorporate piece-wise linear cost curves to model power rating influences.} 

SoC-dependent energy storage models have been widely investigated by experiments and implemented in energy storage \textit{control} models. Previous work~\cite{ecker2014calendar,preger2020degradation,bank2022state} investigated the influence of SoC range and cycle depth on the aging of different Li-ion batteries and found that battery degradation strongly depends on SoC, especially cycling in high and low SoC ranges. Xu et al.~\cite{xu2017factoring, xu2018optimal} incorporated Rainflow cycle depth models in power system optimizations and frequency controls but did not consider the dependency on SoC. Koller et al. \cite{koller2013defining} modeled high and low SoC as a stress factor of battery degradation in model predictive control. The high degradation rate jeopardizes the cycling life of the battery, which leads to higher production (discharge) costs in high and low SoC.  Pandžić~\cite{pandvzic2018accurate} obtained the dependency of battery power rating on SoC and formulated battery power rating using a piece-wise linear approximation in a battery operation model. Yang et al.~\cite{yang2022multi} proposed a multi-state control strategy that determines power rating based on the states of the SoC and the system frequency to enhance the system frequency stability. Zheng et al.~\cite{zheng2015study} examined the correlations between SoC and Coulombic efficiency due to voltage and resistance dependent on SoC. Jafari et al.~\cite{jafari2019improved} investigated a vanadium redox flow battery and modeled its dynamic efficiency and power limits as a function of SoC in a price arbitrage optimization problem. \rev{Motivated by previous studies, we use SoC-dependent energy storage physical models in \textit{market} model comparisons, which is rarely studied in the literature.}

\subsection{\rev{State-of-Charge Management using Market Design}}
\rev{Researchers and electricity market operators are actively seeking new market mechanisms to efficiently manage SoC to better model storage's opportunity cost and physical characteristics~\cite{caiso2020esproposal}. Bhattacharjee et al.~\cite{bhattacharjee2022energy} used a bi-level stochastic optimization model, investigated the implications of different energy storage SoC management entity settings, and found that energy storage SoC self-management could be inefficient under uncertainty. Fang et al.~\cite{fang2022efficient} proposed a bidding structure and a corresponding clearing model for energy storage integration in the day-ahead market. The proposed advanced Vickery-Clarke-Groves mechanism incentivizes energy storage participants to truthfully submit parameters and economically manage SoC by incorporating end-period SoC value. Chen and Tong~\cite{chen2022convexifying} examined energy storage wholesale market participation using convexified bids under a multi-period economic dispatch setting. }

\rev{Compared to previous works that seek to better manage storage SoC with existing market designs, we propose to directly incorporate SoC-dependency in market models in day-ahead and real-time markets to explore the impact of market models on energy storage arbitrage profits and market efficiencies. % To incorporate SoC-dependency in energy storage models, we model these SoC-dependent parameters using piece-wise linear approximations and engineer them into energy storage bids using analytical dynamic programming. 
}

\section{Formulation}\label{form}

We first define an SoC-dependent energy storage model in which the power rating, efficiency, and discharge cost depend on storage SoC. We describe how to incorporate SoC-dependent energy storage model into multi-period optimizations, which we will use as a benchmark for comparison. We then present the single-period SoC segment market model, which dispatches storage using SoC-dependent charge and discharge bids and SoC-dependent energy storage model.

\subsection{State-of-charge Dependent Energy Storage Model}

\rev{We consider a generalized piece-wise linear storage model in which the storage has $S$ segments of discharge cost, efficiency, and power rating parameters to model their dependency over SoC, as illustrated in Section~\ref{socdep}.} Each segment $s \in \mathcal{S} = \in\{1,\dotsc, S\}$ is specified according to the following parameters:
\begin{itemize}
    \item $E_s$ denotes the ending SoC range of segment $s$ in MWh. Segment $s$ has an SoC range between $E_{s-1}$ and $E_s$, $E_0$ is the lower SoC limit, and $E_S$ is the upper SoC limit.  The sequence of the SoC segment index $s$ is monotonic with the SoC range, i.e., $E_s > E_i$ for all $i<s$.
    \item $C_s$ is the physical discharge cost of segment $s$, in \$/MWh; this cost includes prorated cost based on battery degradation and other maintenance costs, but does not cover the electricity cost of charging the storage.
    \item $D_s$ is the maximum discharge power output of segment $s$ standardized with the dispatch time step, in MWh.
    \item $P_s$ is the maximum charge power output of segment $s$ standardized with the dispatch time step, in MWh.
    \item $\eta\up{d}_s$ is the discharge efficiency of segment $s$, $\eta\up{d}_s \in [0,1]$.
    \item $\eta\up{p}_s$ is the charge efficiency of segment $s$, $\eta\up{p}_s \in [0,1]$.
\end{itemize}

\begin{figure}[b]
    \centering
    \vspace{-0.6cm}
	\subfigtopskip=2pt
	\subfigbottomskip=2pt
	\subfigcapskip=-5pt
    \subfigure[]{
    \includegraphics[width=0.48\columnwidth, trim = 10mm 65mm 10mm 65mm, clip]{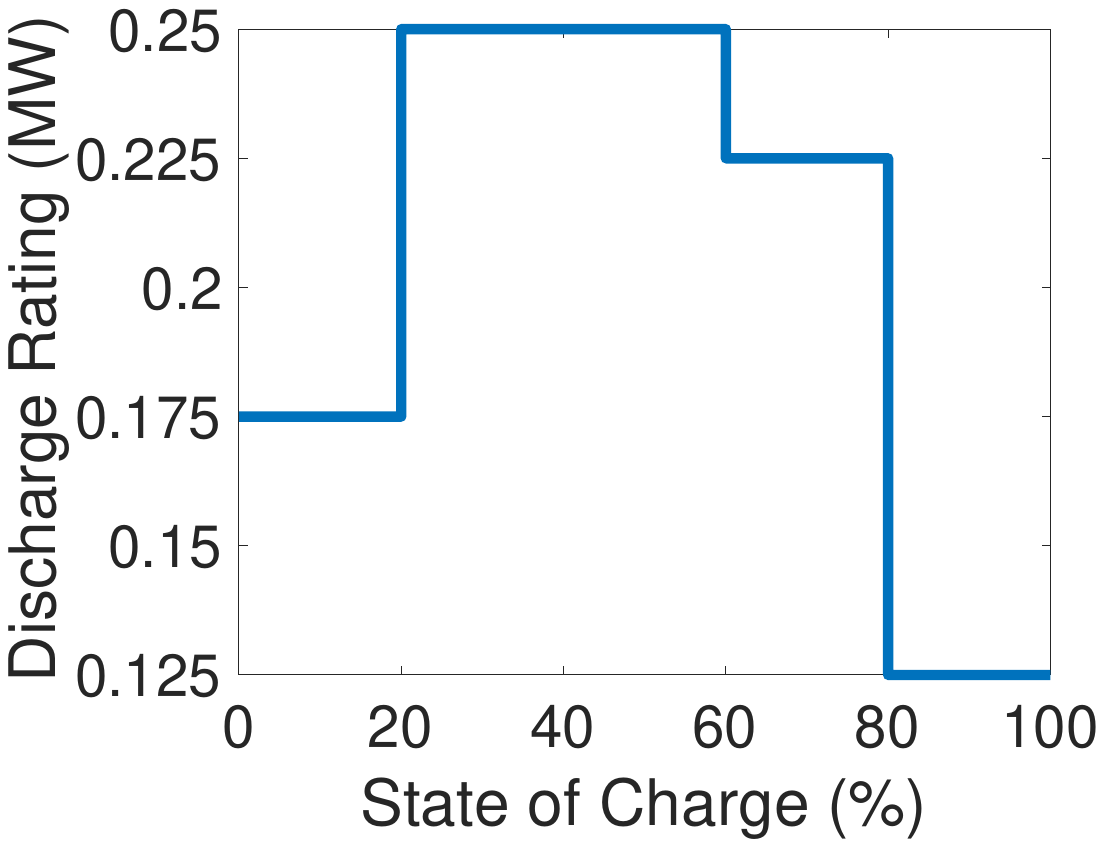}}
    \subfigure[]{
    \includegraphics[width=0.48\columnwidth, trim = 10mm 65mm 10mm 65mm, clip]{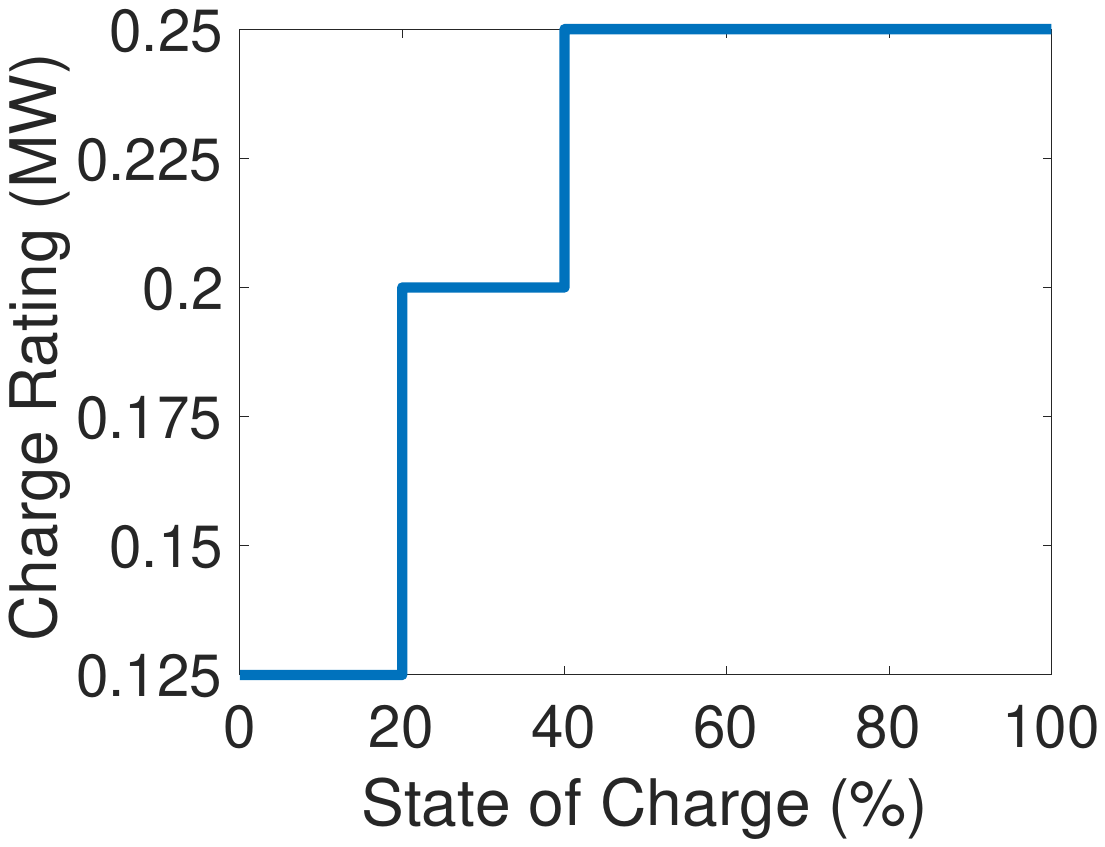}}
    \\
    \subfigure[]{
    \includegraphics[width=0.48\columnwidth, trim = 10mm 65mm 10mm 65mm, clip]{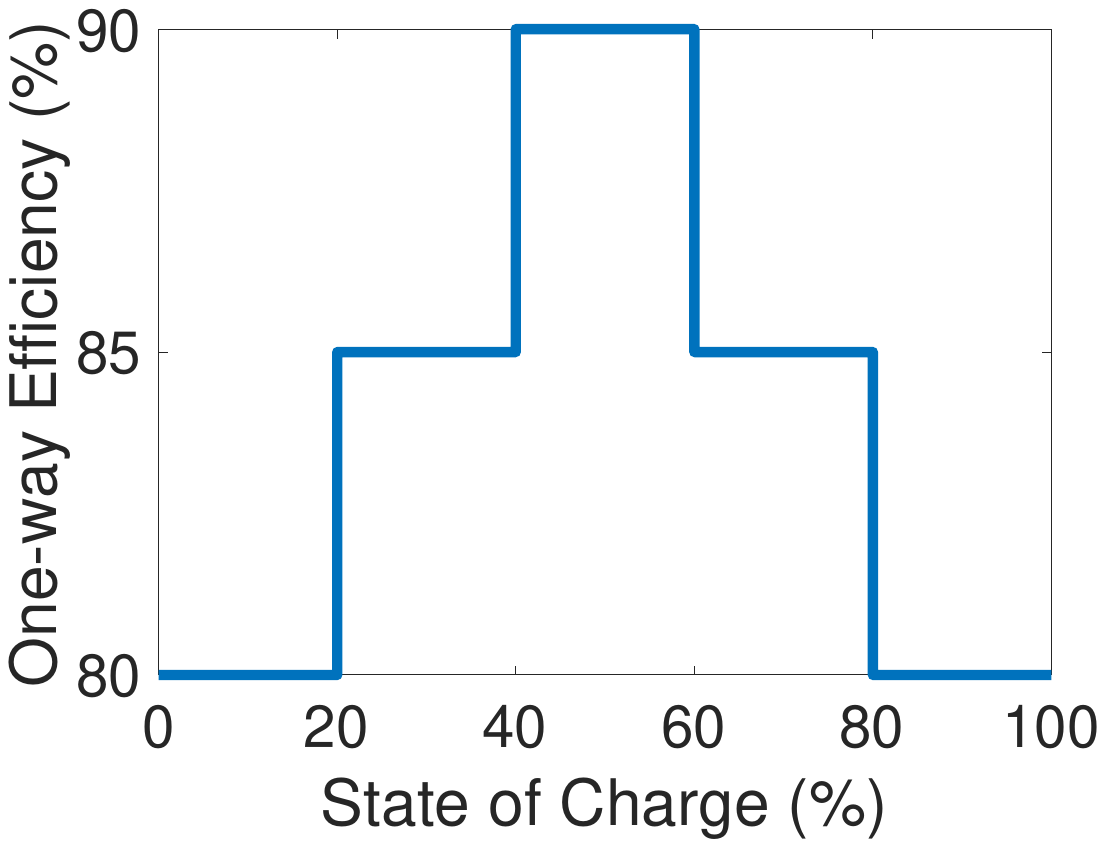}}\label{fig:profit_north}
    \subfigure[]{
    \includegraphics[width=0.48\columnwidth, trim = 10mm 65mm 10mm 65mm, clip]{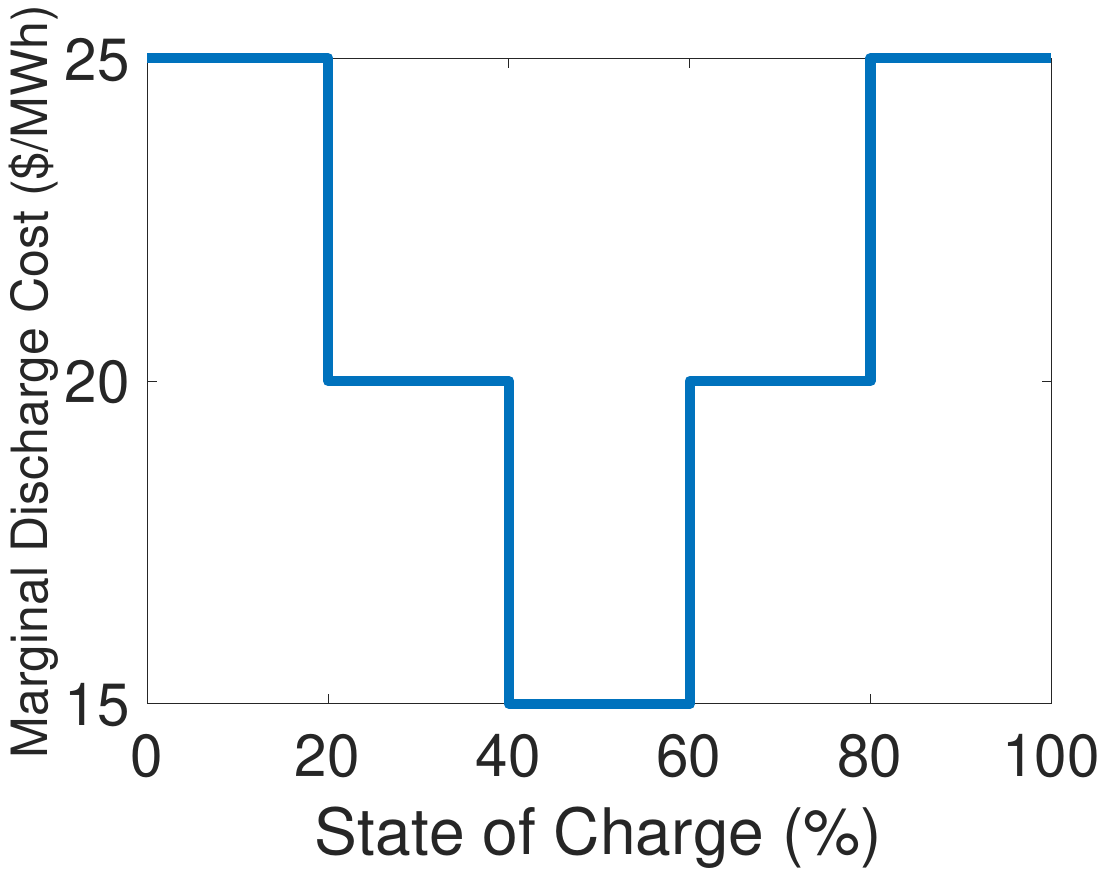}}
    \caption{A 5-segment SoC-dependent energy storage model: (a) discharge rating; (b) charge rating; (c)one-way efficiency; (d) marginal discharge cost.}
    \label{Fig.setting}
\end{figure}

\rev{Fig.~\ref{Fig.setting} shows a 5-segment example of the storage model parameters, where charge and discharge efficiency are simplified to be identical but we can model them as different parameters. } 

We present a multi-period model that dispatches storage directly using SoC-dependent parameters. Although in theory power system operators can use multi-period models to dispatch energy storage in real-time using a look-ahead dispatch setting, this only applies to dispatch decisions from the first time period and repeats the dispatch optimization with updated system states and prediction horizon. As discussed in Section~\ref{sec:2a}, the multi-period dispatch may not be practical due to computation and prediction complexities. Therefore, the primary motivation for using the multi-period dispatch model is to provide a benchmark for comparison with the single-period dispatch model.

We consider a generalized multi-period energy storage dispatch model with a time-varying operating cost function $J_t(\cdot)$ over a time horizon of $t\in\mathcal{T} = \{1,2,\dotsc, T\}$. We introduce binary variables to enforce the SoC segment transition logic. We formulate the dispatch problem as
\begin{subequations}\label{p1}
\begin{gather}
\min_{p_{t,s},d_{t,s}} \; \sum_{t\in\mathcal{T}} J_t\Big(\sum_{s\in\mathcal{S}} (p_{t,s}-d_{t,s})\Big) + \sum_{s\in\mathcal{S}} C_s d_{t,s} \label{p1_obj}\\
\textbf{s.t.}\; 
% 0 \leq p_{t,s} \leq P_s,\; 0\leq d_{t,s} \leq D_s \label{p1_c1}\\
\rev{\sum_{s\in\mathcal{S}} (p_{t,s}/P_s) \leq v_t, \sum_{s\in\mathcal{S}} (d_{t,s}/D_s) \leq 1-v_t,\forall t\in\mathcal{T}} \label{p1_c1}\\
% p_{t,2} \leq  P_2(1-p_1/P_1)\\ % p_1 < P_1 = 0.7, P_2 = 1
% p_{t,s} \leq  P_s(1-\sum_{s-1} p_s/P_s)\\ % p_1 < P_1 = 0.7, P_2 = 1
% 1 hour resolution, charge, 1st segment 0.7MW/1MWh, 2st 1MW/1MWh, starting soc is 0.35MWh, assume 100% efficiency
% \sum_s p_{t,s} \leq k\sum_{s} e_{t,s} + b
e_{t,s} - e_{t-1,s} = -d_{t,s}/\eta\up{d}_s + p_{t,s}\eta\up{p}_s ,\;\forall t\in\mathcal{T}, s\in\mathcal{S}\label{p1_c2}\\
% E_1u_{t,1} \leq e_{t,1} \leq E_1 \label{p1_c3}\\
(E_s - E_{s-1})u_{t,s} \leq e_{t,s} \leq (E_s - E_{s-1})u_{t,s-1},\nonumber\\ \;\forall t\in\mathcal{T}, s\in\mathcal{S}. \label{p1_c3}% , \quad \forall s\in\{2,...,S-1\} \label{p1_c4}\\
% 0 \leq e_{t,S} \leq E_Su_{t,S-1}\label{p1_c5}
\end{gather}\label{MultiModel}

The decision variables in \eqref{MultiModel} include non-negative continuous variables  $p_{t,s}, d_{t,s}$, $e_{t,s}$ and binary variables $v_{t}$ and $u_{t,s}$.
$p_{t,s}, d_{t,s}$ are the battery charging/discharging power output over time period $t$ from SoC segment $s$, and $e_{t,s}$ is the state of energy stored in segment $s$ at the end of period $t$. \eqref{p1_c1} enforces the segment-wise power rating for charge and discharge power; the binary variable ensures that the storage cannot charge and discharge simultaneously. \eqref{p1_c2} models the segment energy evolution subject to the segment charge and discharge efficiency. Finally, \eqref{p1_c3} models the SoC segment logic, that a segment of higher SoC levels must be empty if the lower SoC segment is not full; this logic is achieved using binary variables $u_{t,s}$. $u_{t,s}=1$ indicates SoC segment $s$ is full during time period $t$ ($e_{t,s}=E_s-E_{s-1}$), and $u_{t,s}=0$ indicates the segment is not full. The upper energy limit of SoC segment $s$ is limited by $u_{t,s-1}$, forcing $e_{t,s}$ to be zero if the lower SoC segment $s-1$ is not full, i.e., $u_{t,s-1}=0$.

Total power output and storage SoC is simply the sum of each segment, i.e.,
\begin{align}
    p_t = \sum_{s} p_{t,s}\,,\quad 
    d_t = \sum_{s} d_{t,s}\,,\quad 
    e_t = \sum_{s} e_{t,s} 
\end{align}
where $p_t$ is the storage charge power output over time period $t$, $d_t$ is the discharge power output, and $e_t$ is the storage SoC. 
\end{subequations}

\begin{remark}\emph{Binary variables in multi-period dispatch.}
We employ binary variables $u_{t,s}$ to enforce the SoC segment transition logic because if we relax $u_{t,s}$, the SoC segment transition may not follow the correct charging or discharging order. For example, assume we have a 2-segment SoC model. The model should always discharge the upper segment first or charge the lower segment first.
Now assume the lower segment has a lower discharge cost and the battery is fully charged. The optimization will discharge the lower SoC segment first due to the lower discharge cost, which is against the SoC transition logic. On the other hand, assume a 2-segment empty battery where the upper segment has a cheaper discharge cost. In this case, the optimization will first charge the upper segment because it can discharge later at a lower cost, which also violates the SoC transition logic. Therefore, we use binary variables to enforce the SoC transition logic in this generalized storage SoC market segment model.
\end{remark}

\begin{remark}\emph{SoC-independent storage model.} 
If the storage has constant parameters, i.e., a linear SoC independent storage model, it is equivalent to a 1-segment model in which $s\in\{1\}$.
\end{remark}

\subsection{Single-period Bid-based SoC Segment Market Model}
\begin{figure}[t]
    \centering
    \includegraphics[width=0.9\columnwidth, trim = 0mm 0mm 0mm 0mm, clip]{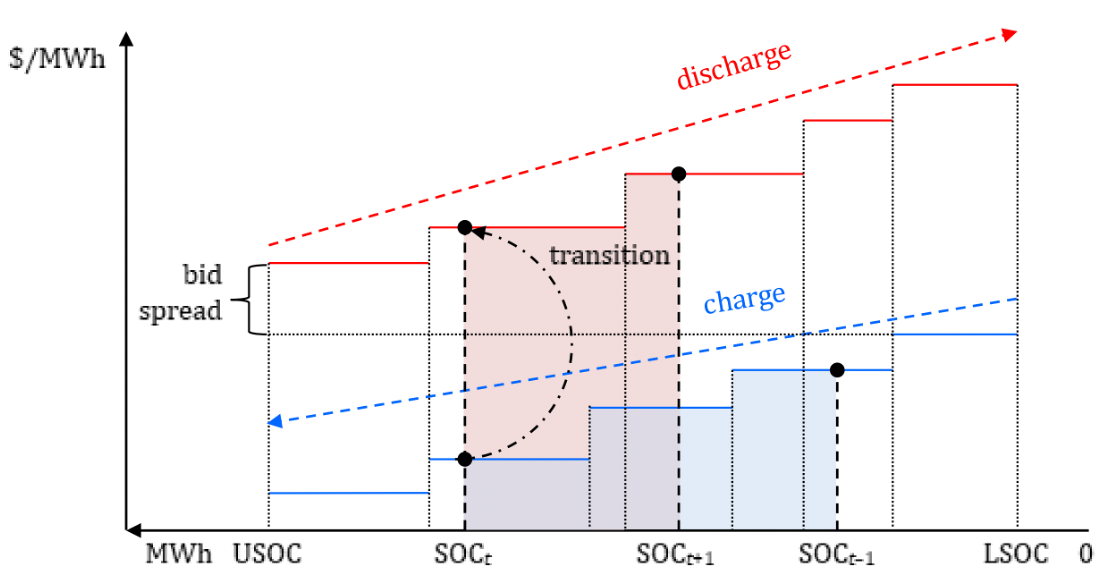}
    \caption{Illustration of SoC segment market model. Charge and discharge bids are segments dependent on SoC ranges.}
    \label{Fig.soc_model}
    \vspace{-1em}
\end{figure}
We now formulate the single-period economic dispatch problem in which the system operator dispatches energy storage based on their submitted bids with physical parameters instead of physical parameters only. We assume the storage submits two sets of time-variant bids, each with $S$ segments. Same as the multi-period model, each bid and parameter segment is associated with an SoC range $E_{s-1}$ to $E_s$.
\begin{itemize}
    \item $G_{t,s}$ is the discharge bid over segment $s$, above which the storage is willing to discharge over segment $s$.
    \item $B_{t,s}$ is the charge bid over segment $s$, below which the storage is willing to charge up over segment $s$.
\end{itemize}
The single-period dispatch model with storage charge and discharge bids is
\begin{subequations}\label{p2}
\begin{gather}
\min_{p_{t,s},d_{t,s}} \quad J_t\Big(\sum_s^S(p_{t,s}-d_{t,s})\Big) + \sum_s^S( G_{t,s}d_{t,s} - B_{t,s}p_s) \label{p2_obj}\\
\textbf{s.t.} \quad (\ref{p1_c1})-(\ref{p1_c3}) \label{p2_c}
\end{gather}\label{SingleModel}which is subject to the same storage constraints as in the multi-period optimization case, but covers only one time period. \rev{We illustrate the proposed SoC segment market model in Fig.~\ref{Fig.soc_model}, where energy storage participants can submit multiple charge and discharge bids between usable SoC (USOC) and lowest SoC (LSOC), depending on SoC segments. We can adjust USOC and LSOC according to the ancillary service award when considering the ancillary services market. The single-period charge and discharge revenues are shown by blue and red shadow areas, respectively.}  If $G_{t,s}$ and $B_{t,s}$ are strictly decreasing within time step $t$, we can relax the problem to linear programming by omitting binary variables.
\end{subequations}

\begin{remark}\emph{Binary variables in single-period dispatch.}
In single-period dispatch, we can remove the binary variable $u_{t,s}$ if $G_{t,s}$ and $B_{t,s}$ monotonically decrease with storage SoC increases. The major difference here is that the optimization considers only a single time period at a time. If we relax binary variables, the single-period dispatch will always try to discharge the segment with the lowest $G_{t,s}$ or charge segments with the highest $B_{t,s}$. If bids monotonically decrease with SoC increases, this becomes equivalent to always discharging SoC segments in descending order and charging in ascending order, which follows the SoC transition logic. 
\end{remark}

\section{Bidding with SoC Segment Market Model}\label{bidmodel}

\rev{We include a bidding model in this paper with a focus on benchmarking the performance of the proposed and existing storage market models in terms of system cost savings, price volatilities, and revenues. While market participants are free to use any approaches to design bids following the proposed market model, the bidding algorithm adopted in this paper is derived theoretically based on dynamic programming and is proven to provide optimal arbitrage decisions~\cite{zheng2022comparing, xu2020operational}. Thus, the comparison reasonably represents results in a competitive electricity market. }

We consider storage participants who design their bids using a profit-maximization price arbitrage model based on a set of price predictions $\lambda_t$. To handle the SoC dependencies in the storage model, we approximate all storage parameters to depend on the storage SoC at the beginning of the time period. The resulting dynamic programming arbitrage problem becomes
\begin{subequations}\label{eq1}
\begin{align}
    Q_{t-1}(e_{t-1}) &= \max_{p_t, d_t} \lambda_t (d_t-p_t) - c(e_{t-1}) d_t + Q_{t}(e_{t}) 
    \label{eq:obj2}
\end{align}
subjects to the following constraints
\begin{gather}
    0 \leq d_t \leq D(e_{t-1}),\; 0\leq p_t \leq P(e_{t-1}) \label{p5_c1} \\
    \text{$d_t = 0$ if $\lambda_t < 0$} \label{p5_c2}\\
    e_t - e_{t-1} = -d_t/\eta\up{d}(e_{t-1}) + p_t\eta\up{p}(e_{t-1}) \label{p5_c3}\\
    0 \leq e_t \leq E \label{p5_c4}
\end{gather}
\end{subequations}
where $Q_{t-1}$ is the maximized energy storage arbitrage profit dependent on the energy storage SoC at the end of the previous time period $e_{t-1}$. $Q_t$ represents the opportunity value of the energy storage SoC $e_t$ at the end of time period $t$, hence the value-to-go function in dynamic programming. \rev{Note that to model the dependencies of the physical parameters illustrated in Section~\ref{socdep}, here the storage discharge cost $c$, power ratings $D$ and $P$, and efficiencies $\eta\up{d}$ and $\eta\up{p}$ are all functions of $e_{t-1}$.} Hence, in this dynamic programming formulation, we apply local linearization to assume the storage parameters are constant within a single time step. In this case, we can up-sample the time step; for example, instead of solving the problem using a market-clearing time frequency such as 5 minutes, we solve the problem at a 1-minute resolution to obtain a more accurate local linearization to the nonlinear storage model. 

We now design bids based on factoring the marginal discharge cost and charge value counting in both physical and opportunity costs
% In the objective function, the first term $\lambda_t (d_t-p_t)$ represents market income, and the rest represents the cost to operate the storage: $c$ represents the physical discharge (mainly due to degradation) cost, and $Q_{e_t}$ represents the opportunity cost (opportunity value remaining). 

\begin{subequations}
\begin{align}
    \partial\frac{c(e_{t-1})d_t - Q_t(e_t)}{\partial d_{t}} &= c(e_{t-1}) + \frac{1}{\eta\up{d}_s}q_{t}(e_t) \\
    \partial\frac{c(e_{t-1})d_t - Q_t(e_t)}{\partial p_{t}} &=  -\eta\up{p}_sq_{t}(e_t)
\end{align}
where $q_t$ is the derivative of $Q_t$. To generate bids for each $G_{t,s}$ and $B_{t,s}$  segment, we replace the discharge cost $c$ with the segment discharge cost $c_s$. For $q$, we take its average value between the SoC range $E_{s-1}$ to $E_s$ by sampling SoC
\begin{align}
    G_{t,s} &\approx  c_s + \frac{1}{\eta\up{d}_s} \sum_{i} \frac{q_t(e_{i,s})}{N_s} \\
    B_{t,s} &\approx   \eta\up{p}_s \sum_{i} \frac{q_t(e_{i,s})}{N_s} 
\end{align}
where $e_{i,s}\in [E_{s-1}, E_s]$ is the SoC samples and $N_s$ is the number of samples. 
% In our prior works~\cite{zheng2022comparing, xu2020operational} we have developed a computation-efficient algorithm to solve the dynamic programming problem so we can obtain the value of $q_t$ over the entire SoC range,  Appendix~\ref{app:q} shows the algorithm.

\end{subequations}
% where $q_t$ is the derivative of $Q_t$, $\mathrm{mean}\big(q(e)\big)|_{E_{s-1}}^{E_s} $ represents the taking the average value of $q_t(e)$ with $e$ between the range of $E_{s-1}$ to $E_s$. In our prior works~\cite{zheng2022comparing, xu2020operational} we have developed a computation-efficient algorithm to solve the dynamic programming problem so we can obtain the value of $q_t$ over the entire SoC range,  Appendix~\ref{app:q} shows the algorithm.

% We define $q_t$ as the derivative of storage opportunity value function $Q_t$ in \eqref{eq:obj2}, which represents the marginal opportunity value of energy stored in the storage. It is evidently that $Q_t$ is valued and differentiable over the energy storage SoC level $e_t$ within [0\,,\,$E$]. We can move the derivative operation into the expectation calculation in \eqref{eq:obj2}. Then we can use an analytical formulation to calculate the opportunity value $q_t(e)$ at any given energy storage SoC level.

Our prior work~\cite{xu2020operational} proposed an analytical algorithm that calculates $q_{t}$ recursively in reverse order. This algorithm uses the following equation to calculate the value $q_{t-1}$ to an SoC input $e$ based on the opportunity function from the next time period $q_t$ as
% can be recursively calculated with next period value function $q_t$, power rating $P$, and efficiency $\eta^d/\eta^p$. We rewrite this value function calculating using a deterministic formulation investigated in this paper as
\begin{align}\label{eq3}
    &q_{t-1}(e) = \nonumber\\
    &\begin{cases}
    q_{t}(e+P\eta^p)  & \text{if $\lambda_{t}\leq q_{t}(e+P\eta^p)\eta^p$} \\
    \lambda_{t}/\eta^p  & \text{if $ q_{t}(e+P\eta^p)\eta^p < \lambda_{t} \leq q_{t}(e)\eta^p$} \\
    q_{t}(e) & \text{if $ q_{t}(e)\eta^p < \lambda_{t} \leq [q_{t}(e)/\eta^d + c]^+$} \\
    (\lambda_{t}-c)\eta^d & \text{if $ [q_{t}(e)/\eta^d + c]^+ < \lambda_{t}$} \\
    & \quad\text{$ \leq [q_{t}(e-D/\eta^d)/\eta^d + c]^+$} \\
    q_{t}(e-D/\eta^d) & \text{if $\lambda_{t} > [q_{t}(e-D/\eta^d)/\eta^d + c]^+$}.
    \end{cases}
\end{align}

Note that in \eqref{eq3}, $c$, $\eta^p$, $\eta^d$, $P$, $D$ are all functions of $e$, but we omitted the function form for simpler presentation.  Since $e$ is input to \eqref{eq3}, the algorithm simply looks up the value of storage parameters based on the input $e$ and finishes the calculation. Thus, we solve the dynamic programming by initializing the final value function $q_T$ as all zeros indicating no more opportunity value at the end of operation horizon, then we perform \eqref{eq3} in reverse order to calculate $q_t$ over the entire time horizon.

We further discretize $q_t$ by equally dividing energy storage SoC level $e$ into small segments, which is finer than the power rating $P$. For any SoC level $e_t$, we can find the nearest segment and return the corresponding value. Note that $Q_t$ in the objective function is the integral of $q_t$. Therefore, discretizing the derivative $q_t$ is equivalent to approximating $Q_t$ using piece-wise linear functions.

\section{Price-taker Case Study}\label{ptcs}
In the first case study, we consider an energy arbitrage problem assuming energy storage is a price-taker with no power to influence market prices. \rev{The objective of the price-taker case study is to \emph{simulate} real-time market clearing with various storage market models while assuming the storage would not impact the market price. In this case, the market clearing problem - in which the system operator minimizes the total system operating cost, can be equivalently modeled as an arbitrage profit-maximizing problem~\cite{castillo2013profit}. However, the key difference between our considered price-taker market clearing model and a price-response/self-schedule model is that the storage’s action is \emph{constrained} by its bids, which must be submitted one hour ahead of time and the bid must stay the same for each hour which contains twelve market clearing time periods. Storage charge and discharge decisions in our study have to be a result of comparing the submitted bids and the market clearing price.
}

To focus on the comparison of market models,
we assume the storage can predict price perfectly but must obey the market rule to design bids and be cleared. This case study aims to demonstrate how different storage market models would impact storage arbitrage profit potential. In terms of market design, we consider three market models:
\begin{enumerate}
    \item \textbf{Multi:} the energy storage is not constrained by the market bidding model and can freely make charge and discharge decisions to arbitrage price differences. This case represents the \emph{best possible} arbitrage results and adopts the optimization \emph{multi-period} dispatch model~\eqref{p1}.
    \item \textbf{RTD-1:} the storage submits one charge bid and one discharge bid for each hour, and the system operator clears the storage bids in \emph{single-period} real-time dispatches. This model represents \emph{existing} storage participation models being used now in most ISOs. This case uses optimization model~\eqref{p2} and sets $S=1$ as it assumes a 1-segment model.
    \item \textbf{RTD-5:} similar to \textbf{RTD-1} except the storage submits 5-segment bids as \emph{proposed} in this paper. We use an equally spaced 20\% SoC  segment range; this case uses optimization model~\eqref{p2} and sets $S=5$.
\end{enumerate}

All three cases use a price arbitrage objective function; the $J_t$ function in the objectives \eqref{p1_obj} and \eqref{p2_obj} becomes the product between price and storage total power:
\begin{align}
    \textstyle J_t\Big(\sum_s^S(p_{t,s}-d_{t,s})\Big) = \lambda_t\Big(\sum_s^S(p_{t,s}-d_{t,s})\Big)
\end{align}
where $\lambda_t$ is the market price.

We use the 2016 CAISO real-time locational marginal prices of one node (WALNUT\_6\_N011) with 5-minute resolution. The energy storage submits hourly charge/discharge bids, and the real-time market clears every 5 minutes. Hence, in \textbf{Multi}, the storage optimizes charge and discharge decisions every 5 minutes, while in \textbf{RTD-1} and \textbf{RTD-5}, the storage bids are the same through 1 hour, and will be cleared every 5 minutes.

\begin{table}[ht]
\vspace{-1em}
\footnotesize
\caption{Comparison of SoC-independent case and SoC-dependent case}
\centering
\begin{tabular}{ccc}
\hline
\hline
\Tstrut
 & Multi-segment bidding & Multi-segment parameters \\
 \hline
SoC-independent & \checkmark & \ding{55}   \\

SoC-dependent     & \checkmark & \checkmark  \\     
\hline
\hline
\end{tabular}
\label{tab:case}
\end{table}

\rev{We consider an SoC-independent storage model and several SoC-dependent storage models in this case study to demonstrate the effectiveness of the proposed SoC market model as shown in Table.~\ref{tab:case}. In both case we apply multi-period dispatch model (\textbf{Multi}) and single-period dispatch model with SoC segment market model (\textbf{RTD-1} and \textbf{RTD-5}). We assume a 4-hour battery energy storage with 1~MWh capacity in all models. In the SoC-independent storage model, we use constant parameters assume the charge/discharge power rating of 0.25~MW (normalized according to 4-hour energy storage with 1~MWh capacity), one-way charge/discharge efficiency of 90\%, and marginal discharge cost of \$20/MWh for all segments. In contrast, SoC-dependent storage model parameters are approximated to 5 segments of step-wise linear functions as shown in Fig.~\ref{Fig.setting}.}  
% We use CAISO real-time locational marginal prices of one node (WALNUT\_6\_N011) with 5-min resolution in segment bids calculation and energy market dispatch models. The energy storage can submit hourly charge/discharge bids and the real-time market clear every 5 minutes.  
We write our code implementation in Julia with JuMP and Gurobi solver, which is available on GitHub\footnote{\url{https://github.com/niklauskun/CAISO_ES_SOC_MODEL}}. We run all numerical simulations on a laptop with an Apple M1 Pro chip and 16~GB memory.

\subsection{SoC-independent storage Model}
\rev{We first compare the three market models using an SoC-independent storage model in which the storage power ratings, efficiencies, and discharge costs are constant. The year-round result of 2016 is shown in Table~\ref{tab:profit}. The single-period model with SoC segment bids (RTD-5) only loses 2.7\% of profit compared to the multi-period model (Multi) due to hourly bidding limitations. The energy storage cannot change bids according to price/opportunity cost variation within hours and submits averaged bids to the system operator instead. The single-period model with 1-segment bids (RTD-1) loses 9.6\% more profit than RTD-5. The result shows that RTD-5 improves storage profit in the market even energy storage parameters are SoC-independent because RTD-5 incorporates SoC-dependent opportunity value. The single-period dispatch decouples inter-temporal decision variables, which reduces  problem complexity. Therefore, the solution times for single-period models are lower than for the multi-period model.}

\begin{table}[ht]
\vspace{-0.5cm}
\footnotesize
\caption{Arbitrage simulation results with CAISO WALNUT 2016 price data with SoC-independent storage models}
\centering
% \resizebox{0.48\textwidth}{!}{
\begin{tabular}{ccccccc}
\hline
\hline
\Tstrut
Storage & Market & Revenue   & Cost  & Profit & Profit  & Solution \\
Model        &  Model     &   [\$]        &   [\$]    &   [\$]    &  Ratio [\%]     & Time (s) \\
\hline
\multirow{3}{*}{IDP}
\Tstrut
&Multi&23042&3770&19272&100&51\\
&RTD-5&22856&4100&18757&97.3&6\\
&RTD-1&20606&3699&16908&87.7&3\\
% \hline
% \multirow{3}{*}{Nonlinear}&Multi&20803&3228&17575&-&288\\
% &RTD-5&17328&2707&14621&83.2&14\\
% &RTD-1&12468&2389&10079&57.3&39\\
\hline
\hline
\end{tabular}
% }
\label{tab:profit}
\vspace{-0.6cm}
\end{table}
% First, we focus on energy storage with linear segment parameters. We compare the multi-period model (Multi) to single-period models with 5-segment energy bids (RTD-5) and power bids (RTD-1). The performance of multi-period model with perfect real-time price prediction is a baseline indicates the maximum energy storage profit can be reach. The RTD-1 model represents the existing energy storage market model.  

\subsection{SoC-dependent Storage Model} \label{sec.case_comp}
We now compare market designs using SoC-dependent storage models, in which the storage power rating, efficiency, and discharge costs depend on the SoC range. We let the storage operator design bids using the SoC-dependent storage model in RTD-1 and RTD-5. \rev{Only in RTD-1, the dispatch instructions are not always feasible because the system operator clears the market using an SoC-independent (1-segment) storage model while the actual storage model is SoC-dependent. Energy storage submits average discharge cost, highest efficiency, and highest power rating as industrial implementation to maximize their profit, provided most arbitrage profit comes from sparse abnormal prices. To bound energy storage dispatches within the physically feasible region of the storage in RTD-1, we project the dispatch instruction to the storage feasible operation region as in (\ref{p2_c}) by minimizing the square error}, i.e.:
\begin{align}
    \textstyle \min_{p_{t,s},d_{t,s}} ||\sum_{s}p_{t,s} - \hat{p}_t||^2 + ||\sum_{s}d_{t,s} - \hat{d}_t||^2 \label{p3_obj}\\
    \text{subjects to \eqref{p2_c}}
\end{align}
where $\hat{p}_t$ and $\hat{d}_t$ are the dispatch instructions cleared by the system operator. In this setting,  we simulate the process of SoC-dependent energy storage submitting bids to the system operator and trying their best to follow the received dispatch signals from the system operator.

We design five SoC-dependent storage models with different power ratings, motivated by our observation that SoC-dependent storage power ratings are the primary factor affecting storage profit potentials: 
\begin{enumerate}
    \item \textbf{DPA:} the storage has SoC-dependent segment-wise charge and discharge power ratings as shown in Fig.~\ref{Fig.setting}.
    \item \textbf{DPB:} same as DPA except the storage has a constant discharge power rating.
    \item \textbf{DPC:} same as DPA except the storage has constant charge and discharge ratings.
    \item \textbf{DPF:} the storage has a constant charge power rating but the discharge power ratings decrease in the first segment. Discharge power ratings are [70\%, 100\%, 100\%, 100\%, 100\%] of the highest power rating; otherwise, the model is the same as DPA.
    \item \textbf{DPL:} the storage has a constant charge power rating but the discharge power ratings decrease in the last two segments. The segment-wise discharge power ratings are [100\%, 100\%, 100\%, 90\%, 50\%] of the highest power rating; otherwise, the model is the same as DPA.
\end{enumerate}

\begin{table}[ht]
\vspace{-1em}
\caption{Arbitrage simulation results with CAISO WALNUT 2016 price data with SoC dependent storage models}
\centering
% \resizebox{0.48\textwidth}{!}{
\begin{tabular}{ccccccc}
\hline
\hline
\Tstrut
Storage & Market & Revenue   & Cost  & Profit & Profit  & Solution \\
Model        &  Model     &      [\$]     &   [\$]    &   [\$]    &  Ratio [\%]     & Time (s) \\
\hline
% \multirow{3}{*}{Linear}&Multi&23042&3770&19272&100&53\\
% &RTD-5&22856&4100&18757&97.3&15\\
% &RTD-1&20606&3699&16908&87.7&15\\
% \hline
\multirow{3}{*}{DPA}
\Tstrut
&Multi&20803&3228&17575&100&283\\
&RTD-5&17328&2707&14621&83.2&22\\
&RTD-1&12468&2389&10079&57.3&16\\
\hline
\multirow{3}{*}{DPB}
\Tstrut
&Multi&21233&3148&18085&100&193\\
&RTD-5&20372&2995&17377&96.1&19\\
&RTD-1&17544&2834&14710&81.3&11\\
\hline
\multirow{3}{*}{DPC}
\Tstrut
&Multi&22081&3310&18772&100&97\\
&RTD-5&21450&3382&18068&96.2&18\\
&RTD-1&18529&3056&15472&82.4&11\\
\hline
\multirow{3}{*}{DPF}
\Tstrut
&Multi&21905&3470&18435&100&170\\
&RTD-5&20174&3299&16875&91.5&18\\
&RTD-1&17261&2851&14410&78.2&11\\
\hline
\multirow{3}{*}{DPL}
\Tstrut
&Multi&21833&3236&18598&100&172\\
&RTD-5&19898&3282&16616&89.3&18\\
&RTD-1&14462&2893&11569&62.2&11\\
\hline
\hline
\end{tabular}
% }
\label{tab:nonprofit}
\end{table}

Table~\ref{tab:nonprofit} shows the result for the entire year of 2016. In DPA, RTD-5 gets a lower profit ratio than the SoC-independent storage model case, and RTD-1 only captures 57.3\% profit compared to Multi. However, in DPB, when the discharge rating is constant, we reached a profit ratio close to the SoC-independent case. We further investigate two SoC-dependent energy storage models with different discharge power rating curves (DPL and DPF). The performances of single-period models are better than the DPA case, especially in DPF, which only has reduced power rating at lowest SoC segment. The charge power rating has little effect on the profit ratio, as shown in the DPC case. 

\begin{figure}[t]
    \centering
    \includegraphics[width=0.90\columnwidth,trim = 5mm 0mm 10mm 0mm, clip]{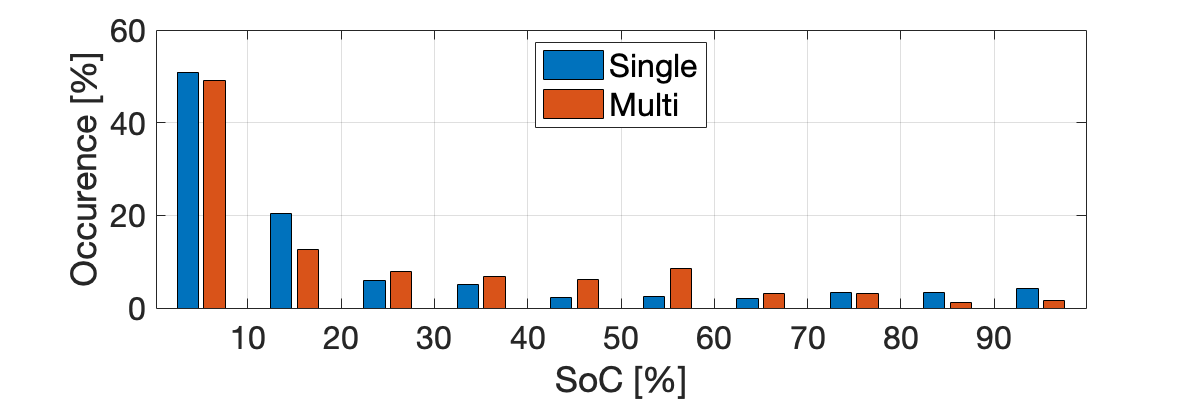}
    \vspace{-0.3cm}
    \caption{Histogram comparing SoC distribution in multi-period (Multi) and single-period dispatch (RTD-5) results of the nonlinear storage model.}
    \label{fig:sochist}
\end{figure}

RTD-5 provides a lower profit, primarily because the storage has to discharge at a lower power rating when price spikes occur, and the SoC is not positioned at the 20\% to 60\% range with the highest power rating. In the single-period dispatch, the storage is more likely to charge up to high SoC values or discharge to low SoC values based on the submitted bids, while the multi-period optimization better optimizes operation based on SoC segments. 
% There are two scenarios that lead to this result. First, the energy storage predicts  latter price spikes, which results in higher charging bids and the storage charge to a higher state-of-charge level compared to the multi-period model result. Second, the total energy required in the price spike duration is not large enough for the energy storage to charge to a high state-of-charge level. In both of these two scenarios, the energy storage is not at optimal discharge segments and can not discharge at full power rating when price spikes realize. 
Fig.~\ref{fig:sochist} shows a histogram comparing SoC distribution in the DPA model with the Multi and RTD-5 cases. The energy storage stays between 20\% to 60\% more often in the Multi case, so the storage can discharge at a higher power rating more frequently than the RTD-5 case.

% \subsection{Sensitivity Analysis}
% We explore the sensitivity of real-time price forecast accuracy and the number of bid segments under the linear energy storage setting in this subsection. We first test charge and discharge bids calculated with different noise levels of real-time price. As shown in Fig.~\ref{fig:noise}, the captured profit decreases as the real-time price forecast loses fidelity, but the performance is robust even with a high noise level. In Fig.~\ref{fig:seg}, the revenue increases and converges as the number of segments increases, but the profit slightly decreases when the number of segments is large enough. Increasing number of bid segments 

% \begin{figure}
%     \centering
%     \subfigure[]{\includegraphics[width=0.48\columnwidth,trim = 10mm 65mm 10mm 65mm, clip]{figure/noise.pdf}}
%     \subfigure[]{\includegraphics[width=0.48\columnwidth,trim = 10mm 65mm 10mm 65mm, clip]{figure/segment.pdf}}\label{fig:seg}
%     \caption{Caption}
% \end{figure}

% \begin{figure}
%     \centering
%     \includegraphics[width=0.9\columnwidth,trim = 10mm 65mm 10mm 65mm, clip]{figure/segment.pdf}
%     \caption{Caption \todo{do one from one to ten segment separated by 1 (1:10)}}
%     \label{fig:seg}
% \end{figure}
\begin{figure*}[htbp]%
	\centering
	\subfigtopskip=2pt
	\subfigbottomskip=2pt
	\subfigcapskip=-5pt
	
	\subfigure[Average operating costs]{
		\includegraphics[trim = 0mm 0mm 0mm 0mm, clip, width = .9\columnwidth]{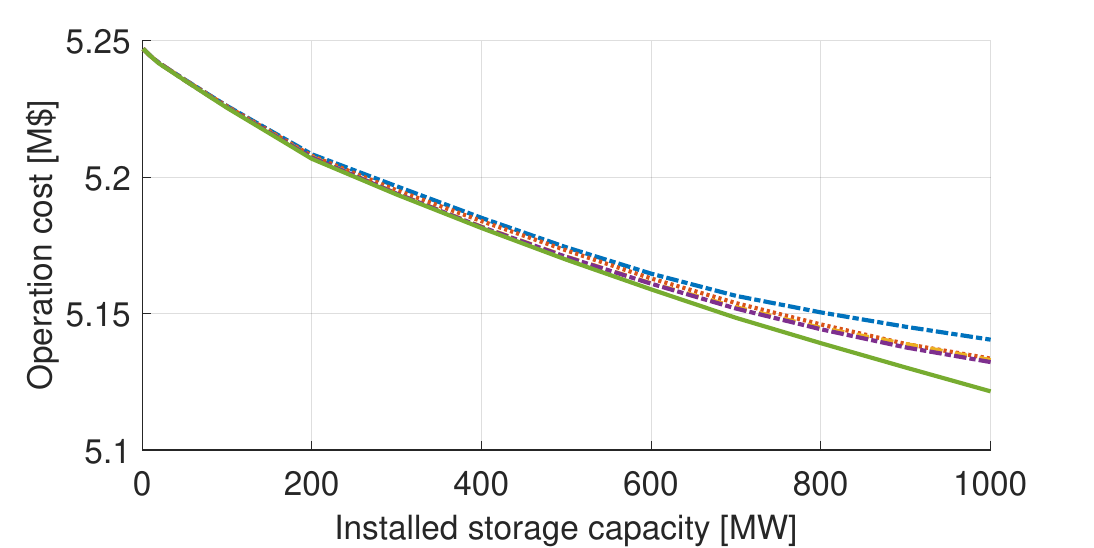}\label{fig:4a}%
	}
	\subfigure[Cost reduction normalized to \textbf{RTD-1}]{
		\includegraphics[trim = 0mm 0mm 0mm 0mm, clip, width = .9\columnwidth]{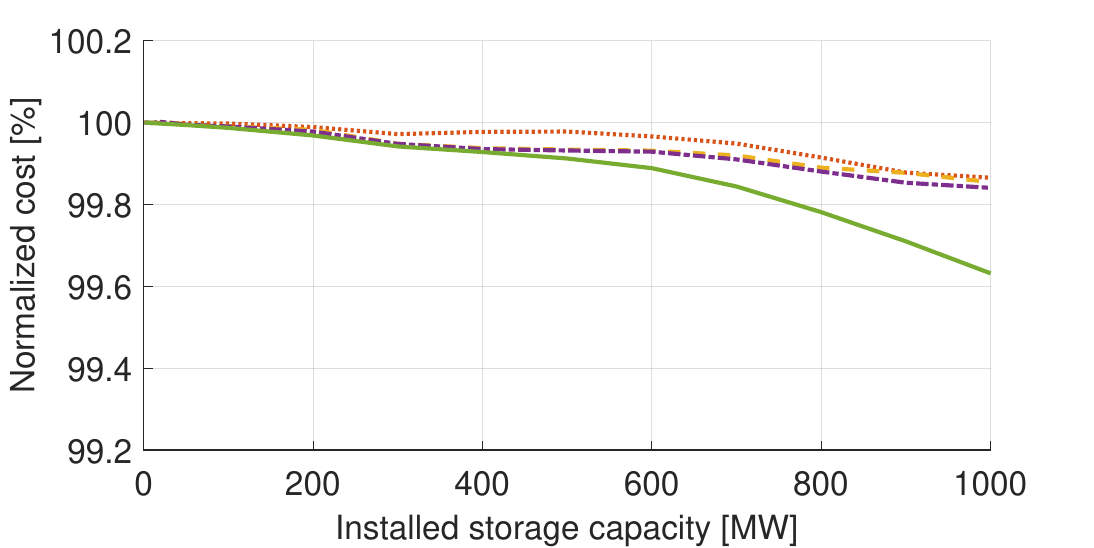}\label{fig:4b}
	}
	\\
	\subfigure[Average real-time prices]{
		\includegraphics[trim = 0mm 0mm 0mm 0mm, clip, width = .9\columnwidth]{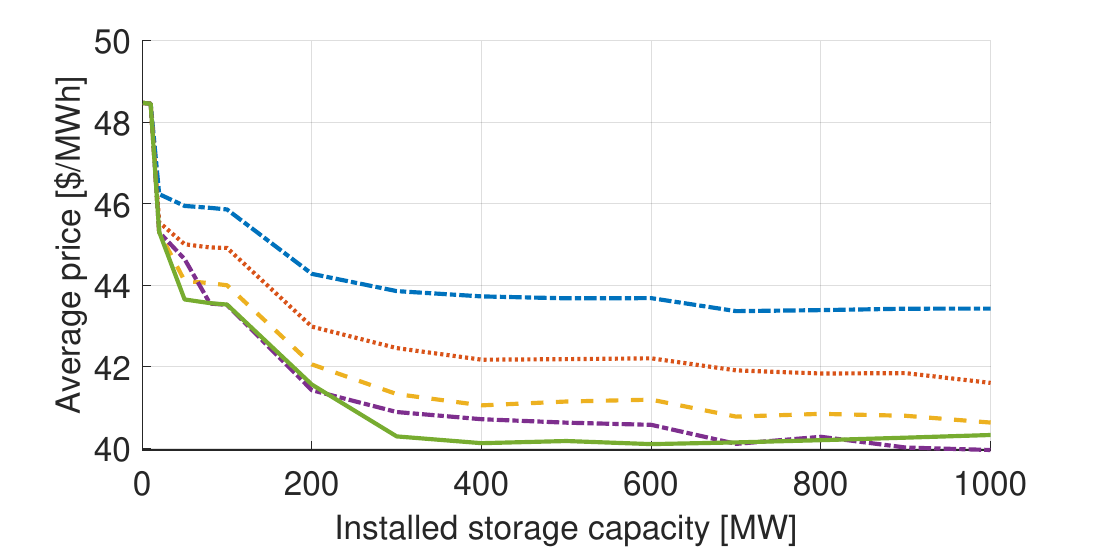}\label{fig:4c}
	}
	\subfigure[Scenario-average price standard deviations]{
		\includegraphics[trim = 0mm 0mm 0mm 0mm, clip, width = .9\columnwidth]{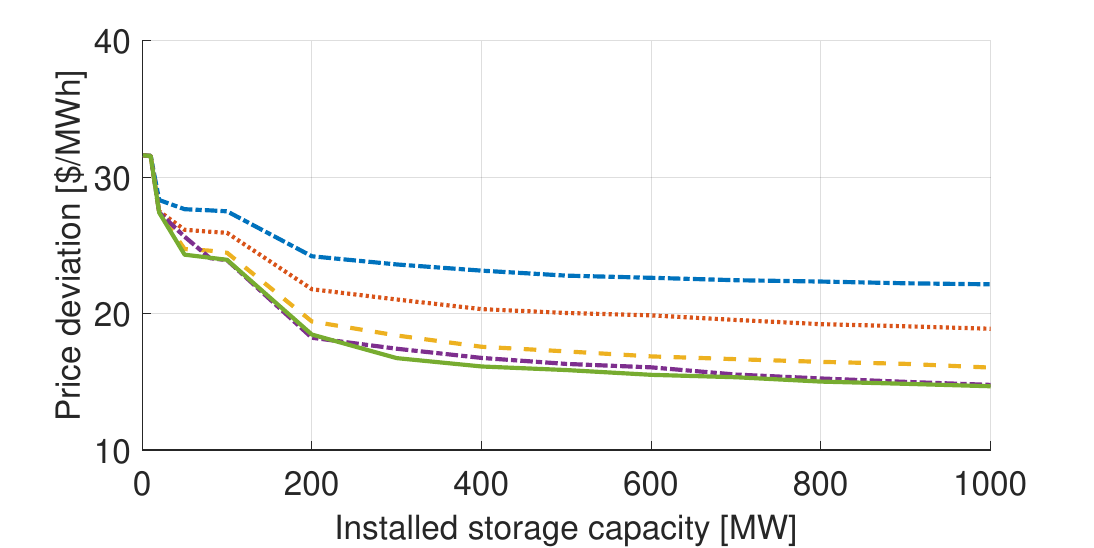}\label{fig:4d}}
	\\
		\subfigure[Average daily profits]{
		\includegraphics[trim = 0mm 0mm 0mm 5mm, clip, width = .9\columnwidth]{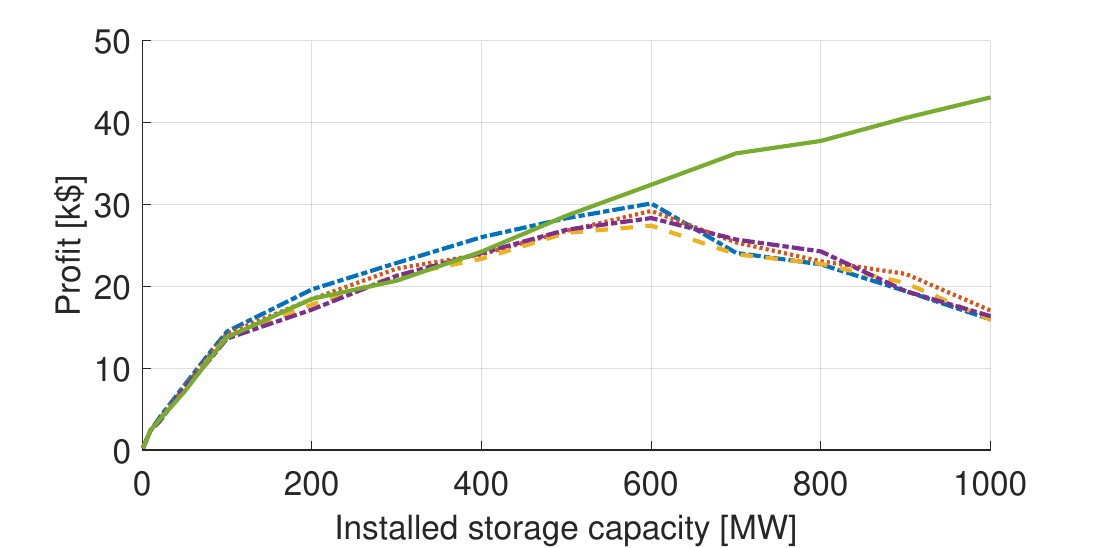}\label{fig:4e}
	}
	\subfigure[Average daily per-MW profits]{
		\includegraphics[trim = 0mm 0mm 0mm 5mm, clip, width = .9\columnwidth]{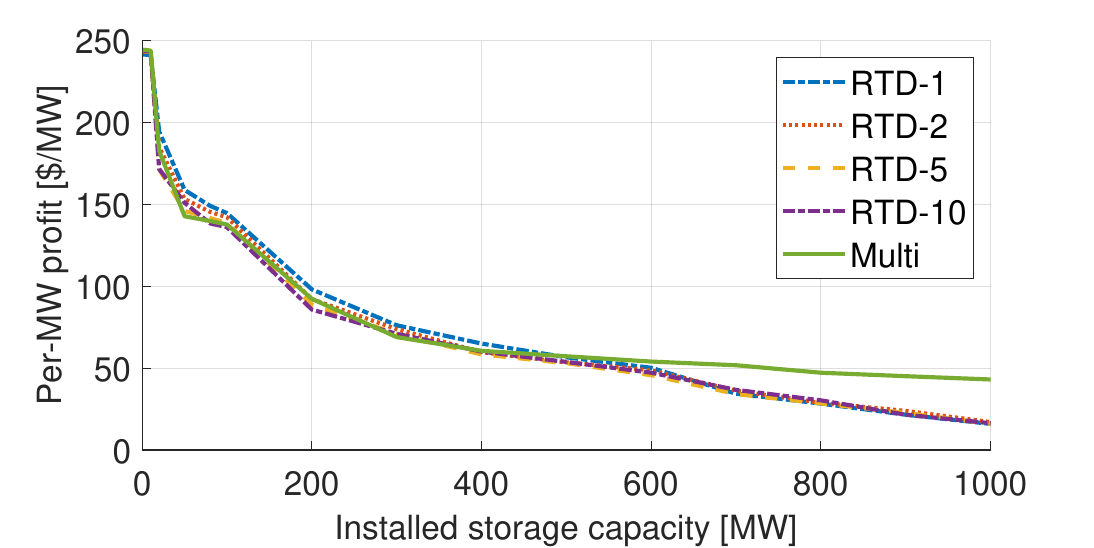}\label{fig:4f}
	}
  \caption{Market clearing results comparison with SoC-independent storage models. \textbf{All figures use the legend in Fig.~\ref{fig:4f}}.}
    \label{fig:compseg}
    \vspace{-1em}
\end{figure*}

\section{Price-influencer Case Study}\label{pics}

\rev{In the second case study, we consider energy storage as a price-influencer, meaning its charge and discharge will affect the price. We assume a competitive market in which the storage participant will not try to exercise market power, but its action will still impact the market prices and other outcomes. 
%We consider two-stage day-ahead and real-time markets. The system operator first clears the day-ahead market over a 24-hour horizon as in \rev{\eqref{gbalance}}, schedules generator commitments, and then performs real-time dispatch with hourly resolutions.
% Since there is no systematic market model to manage storage SoC in the existing real-time market model economically, the case study focuses on evaluating different storage bidding models in a real-time market. 
% \rev{Note that here we only consider storage participating in real-time markets rather than in both markets because it requires new bidding strategies on both day-ahead and real-time, which is beyond the scope of this paper. To assess energy storage real-time market performance in realistic deregulation power system operations, the unit commitment result is needed to generate day-ahead price, which is used as real-time price predictions to generate real-time market bids. Since we focus on comparing market models, we do not consider system uncertainties. Therefore we assume day-ahead and real-time have the same demand and renewable profile, i.e., perfect demand and renewable predictions. The only difference is that there is no storage in the day-ahead market, but storage participates in the real-time market.}
Since the focus of the study is to compare different storage market models, we assume the storage only participates in real-time markets and has a perfect prediction of the day-ahead market prices cleared without energy storage participation, but cannot predict how its market actions will impact the real-time price clearing. To do this, we first perform a day-ahead 24-hour unit commitment in Appendix\ref{app:uc} without storage and record the dispatch cost, commitment status, and prices. The storage will use the price results from the unit commitment and design bids using the considered market rule, and then we perform both a multi-period dispatch and a sequential single-period dispatch using the same demand and wind profile, and the commitment results from unit commitment. Two types of dispatch settings are compared:}
\rev{
\begin{enumerate}
    \item \textbf{Multi-period dispatch (Multi)}: The system operator solves a 24-hour period economic dispatch with storage's physical cost as in Appendix\ref{app:ed}~\eqref{gbalance1}. The Multi-period case also serves as a benchmark as it provides the optimal dispatch result if assuming the same storage model and forecast accuracy.
    % \rev{with different numbers of SoC segments} as in \rev{\eqref{gbalance1}}. \rev{This case aims to maximize social welfare outcome under the given SoC segment, where Multi-10 is a benchmark case with 10 SoC segments to achieve the best social welfare, i.e., lowest operation costs};
    \item \textbf{Single-period dispatch (RTD)}: The system operator solves sequential real-time dispatch problems with storage bids as in Appendix\ref{app:ed}~\eqref{gbalance3}. These bids cover both the physical and opportunity cost of the storage. 
\end{enumerate} }
% In the \textbf{multi-period dispatch}, we solve a 24-hour period economic dispatch with storage physical parameters as in \eqref{p1}; for the \textbf{single-period dispatch}, we solve sequential dispatch problems as in \eqref{p2} with the generated bids, and update the storage SoC before each dispatch period. We perform single-period dispatch models with different SoC segment models.  

We do not consider ramp limits so that the multi-period and single-period dispatch yield the same dispatch and price results without storage, as there are no more inter-temporal constraints. % \done{following sentence removed due to redundancy}

% Appendix~\ref{app:uc} shows the unit commitment formulation and Appendix~\ref{app:ed} shows the economic dispatch formulation for both cases.

We perform simulations on an Independent System Operator New England (ISO-NE) 8-zone test system~\cite{krishnamurthy20158}. \rev{The system demand varies between 9~GW to 17~GW, with an average of 13~GW. The wind capacity is 6.5~GW with an average wind capacity factor (average power output divided by its maximum power capability) of 0.4. We pick five representative demand and wind profiles for our study using a K-means approach.
In all cases, we observe the market clearing results with storage capacity increasing from 1~MW to 1000~MW with a 4-hour duration, simulating increasing storage participation in markets. We assume multiple storage have the same action. The code implementation is written in Matlab 2021b with YALMIP as a socket to connect Gurobi solver and Matlab. We run the numerical simulations on a desktop PC computer with an Intel I7-11700 chip and 64 GB memory.}
% The storage parameters are SoC-dependent with nonlinear storage model with a 4-hour duration and varying power capacities and \$10/MWh discharge cost.The energy storage has 90\% one-way efficiency, and the limitation of charging/discharging power rating is 0.25. 
% We assume multiple storage same action 

% The settings for storage efficiency and power ratings are the same as Section~\ref{sec.case_comp}. 

\subsection{SoC-independent Storage Model}

% We further assume the system operator has a perfect forecast of demand and renewable generations in day-ahead unit commitments, and all generators have no ramp constraints. The goal of these assumptions is to ensure without storage participating in real-time markets, the real-time dispatch will provide exactly the same results as the day-ahead unit commitment, including market-clearing prices and system costs. We assume all generators in the system submit their true costs to the system operator. We now assume the storage design market bids based on day-ahead price results and submit bids into real-time dispatches. Note that we assume the storage is a price-influencer but do not try to exercise market power, therefore in the bid design the storage does not assume its action will change market prices. 

\rev{We first consider an SoC-independent storage model to show the benefit of implementing SoC-segment market models even if the storage physical parameters are independent of SoC. In this case, the system operator can dispatch the storage optimally in multi-period dispatch without implementing SoC-segment models. Yet, the SoC-segment model still provides benefits in real-time dispatch as the storage opportunity cost depends on the SoC. In this case study, we have storage with a one-way efficiency of 90\%, total charging/discharging power ratings are normalized according to installed 4-hour energy storage capacity, and marginal discharging cost at \$20/MWh.}

% \rev{In the SoC-independent model, the system operator does not consider SoC dependencies of the storage parameters but only considers storage multi-segment bidding as shown in Table~\ref{tab:case}. In this case, we have storage with a one-way efficiency of 90\%, total charging/discharging power ratings are normalized according to installed 4-hour energy storage capacity, and marginal charging/discharging cost at \$20/MWh.}

Fig.~\ref{fig:compseg} shows the simulation results with different installed energy storage capacities over multi-period and single-period dispatch using different market models.
% The multi-period dispatch corresponds to the system operator performing a 24-hour real-time dispatch with perfect information, representing the best possible social welfare outcome.
Fig.~\ref{fig:4a} shows the averaged total system operating cost comparison, and \ref{fig:4b} shows the normalized cost comparison based on the single-period dispatch 1-segment (RTD-1) market model results. These results conclude that the multi-segment (RTD-2, 5, 10) market model is better at reducing system operating costs, hence improving social welfare. A higher number of segments can further narrow the gap to optimal social welfare. In Fig.~\ref{fig:4b}, we observe that RTD-2 provides a significant improvement compared to RTD-1, while RTD-5 and RTD-10 produce similar results to RTD-2. 
% The most significant difference occurs at 400~MW storage capacity, while 
\rev{RTD-2 further reduces the system operating cost by an additional 0.1\% of the total system cost, or 5\% more cost reduction led by energy storage (compare cost without energy storage to with 1000~MW energy storage) compared to RTD-1. This equals an annual cost savings of around \$2 million in our tested ISO-NE system.}

% Such an improvement becomes increasingly apparent as storage capacity increases: At 400~MW storage capacity, the multi-segment model reduces 0.1\% of the total system operating cost, but at 1000~MW storage capacity, the multi-segment model can reduce 0.5\% of the total cost. This equals an annual cost savings of around \$2 million and \$10 million in our tested ISO-NE system, respectively. The reason is that single-segment market model approximates storage nonlinear parameters, which can generate infeasible charging/discharging commands for energy storage. As storage capacity increases, the storage power has increasing deviations from the dispatch command. Therefore, additional costs are needed to compensate such a power imbalance in frequency regulation market, resulting in higher total system operating costs. 

% \begin{figure}[H]%
% 	\centering
% 	\subfigure[Average daily profits]{
% 		\includegraphics[trim = 5mm 0mm 3mm 0mm, clip, width = .9\columnwidth]{figure/CompSeg_profit.pdf}%
% 	}
% 	\\
% 	\subfigure[Normalized profit to multi-period case]{
% 		\includegraphics[trim = 4mm 0mm 2mm 0mm, clip, width = .9\columnwidth]{figure/CompSeg_seg_profit.pdf}
% 		%\label{fig:bid3}%
% 	}
%   \caption{Market model storage profits comparison.}
%     \label{fig:compseg_pro}
% \end{figure}

Fig.~\ref{fig:4c} and \ref{fig:4d} show the average price and scenario-averaged price standard deviations. The multi-segment model provides significant improvements in reducing market prices and their volatilities, especially with RTD-10 which achieves price results similar to those of Multi. RTD-10 reduced the average price by around 10\% and the price standard deviation by around 30\% compared to the RTD-1 in the range between 200 and 1000~MW storage capacity. On the other hand, the number of segments did not provide much difference in storage profitability, and per-MW profit decreases as the storage capacity increases, as shown in Fig.~\ref{fig:4e} and \ref{fig:4f}. Notably, at a low storage capacity (below 500~MW), RTD-1 provides the highest profit. This is because RTD-1 kept price volatilities as shown in Fig.~\ref{fig:compseg}~(c), maintaining higher profit than all other cases. Yet as storage capacity increases, storage starts to have more potent influences on market price patterns, and the storage profit stops increasing in single-period cases after 600~MW storage capacity.

\subsection{ \rev{SoC-dependent Storage Model}}
\begin{figure*}[t]%
	\centering
	\subfigtopskip=2pt
	\subfigbottomskip=2pt
	\subfigcapskip=-5pt

	\subfigure[Average operating costs]{
		\includegraphics[trim = 0mm 0mm 0mm 0mm, clip, width = .9\columnwidth]{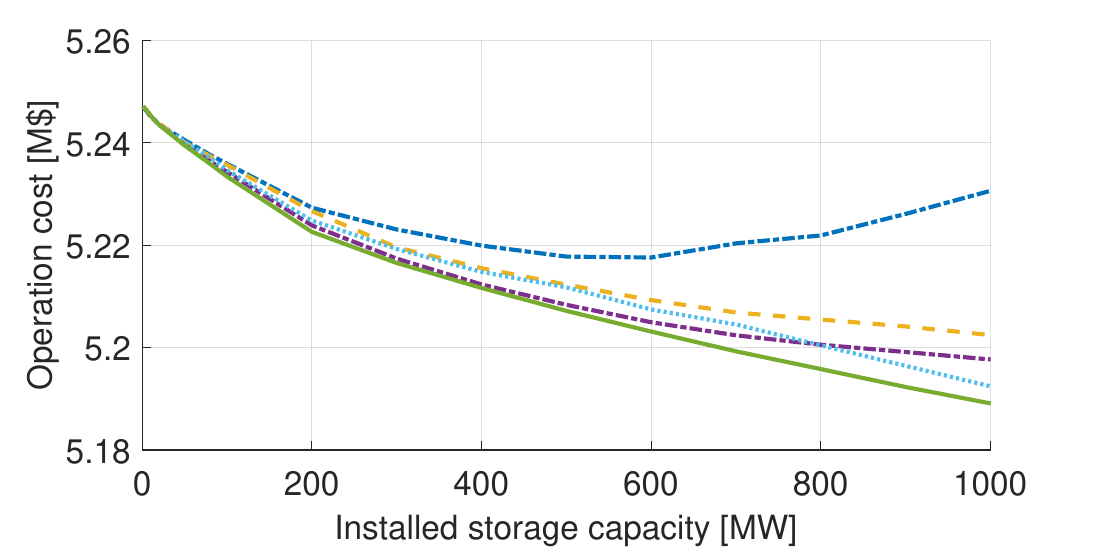}\label{fig:5a}%
	}
	\subfigure[Cost reduction normalized to \textbf{RTD-1}]{
		\includegraphics[trim = 0mm 0mm 0mm 0mm, clip, width = .9\columnwidth]{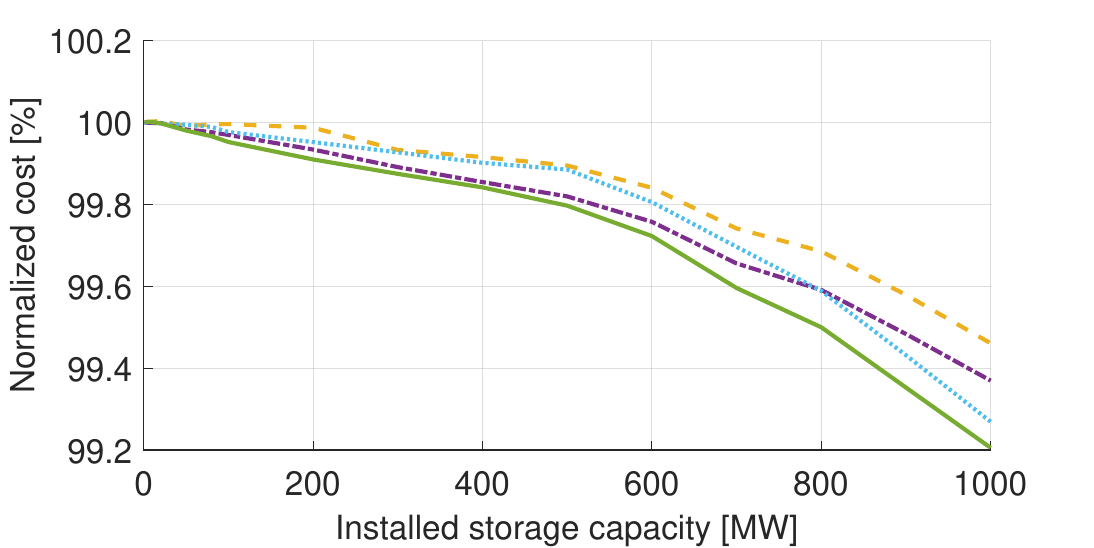}\label{fig:5b}
	}
	\\
	\subfigure[Average real-time prices]{
		\includegraphics[trim = 0mm 0mm 0mm 0mm, clip, width = .9\columnwidth]{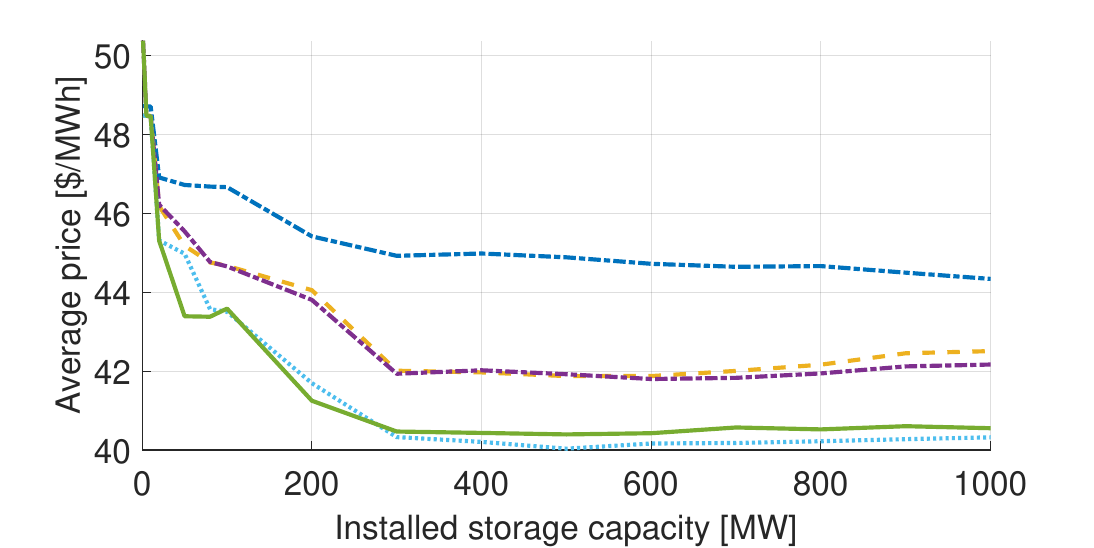}%
	}
	\subfigure[Scenario-average price standard deviations]{
		\includegraphics[trim = 0mm 0mm 0mm 0mm, clip, width = .9\columnwidth]{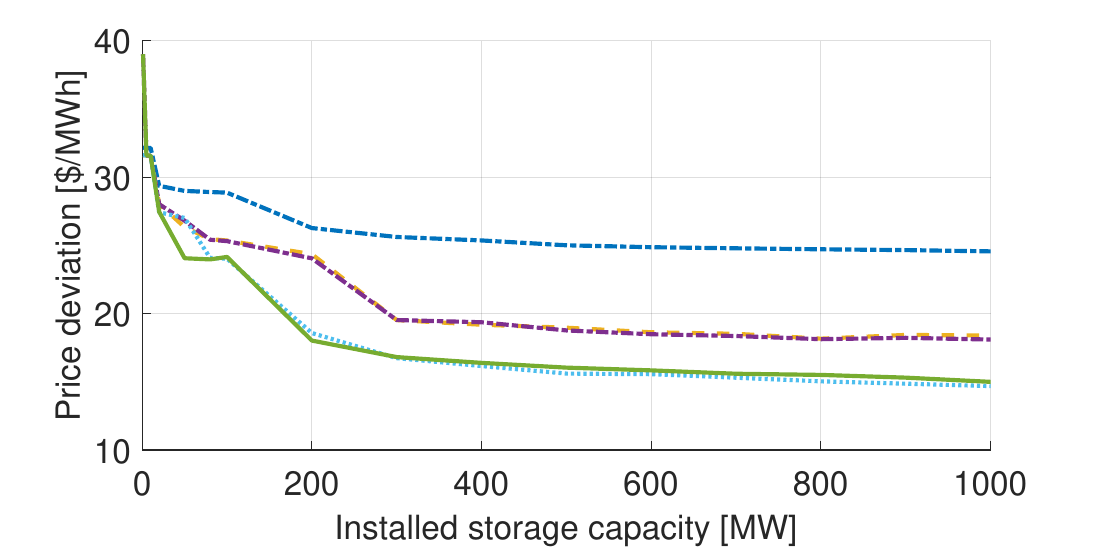}
		%\label{fig:bid3}%
	}
	\\
		\subfigure[Average daily profits]{
		\includegraphics[trim = 0mm 0mm 0mm 0mm, clip, width = .9\columnwidth]{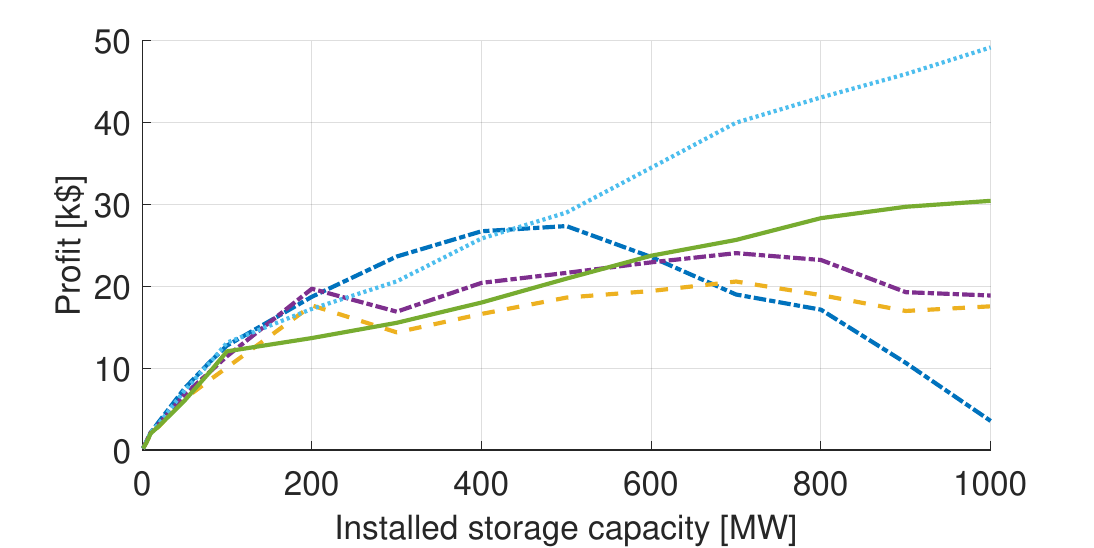}\label{fig:5e}
	}
	\subfigure[Average daily per-MW profits]{
		\includegraphics[trim = 0mm 0mm 0mm 0mm, clip, width = .9\columnwidth]{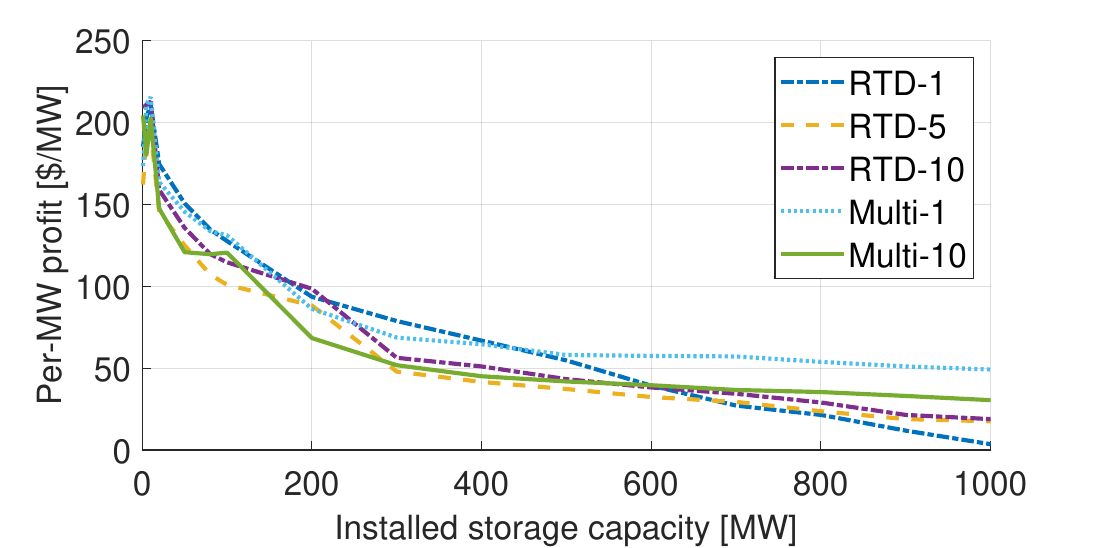}\label{fig:5f}
	}
  \caption{Market clearing results comparison with SoC dependent models. \textbf{All figures use the legend in Fig.~\ref{fig:5f}}.}
    \label{fig:compseg2}
    \vspace{-1em}
\end{figure*}

\rev{We now consider an SoC-dependent model, in which storage physical parameters depend on SoC, to demonstrate the benefit of dispatch storage using SoC-segment models. In this case, modeling SoC segments will improve the dispatch accuracy in both multi-period and single-period dispatch, as an SoC-segment model provides a closer approximation to the true storage model. In the case study, we consider a storage segment model the same as Fig.~\ref{Fig.setting},  prorated to 1, 5, or 10 segments depending on the case study setting. Note that the dispatch may not be feasible in the case of the 1-segment model as we assume the real storage model is 5-segment. Thus, we will again use norm-2 projection as in \eqref{p3_obj}, while the mismatch between the instruction and the actual dispatch is penalized with a \$50/MWh penalty cost.
}
\rev{
We first compare storage market models in RTD. Fig.~\ref{fig:compseg2} shows a similar trend as in the SoC-independent case study that a higher number of SoC segments achieves lower system costs, average price, and price volatility. Yet, the improvement is more significant than the SoC-independent cases because now both the storage physical parameters and opportunity costs depend on SoC. Notably, the 1-segment storage model increased the system cost  at high storage capacity cases, which primarily contributed to the \$50/MWh infeasibility penalty cost. On the other hand, the profit result for storage is mixed, but RTD-1 still provides higher profit at low storage capacity. This result also agrees with the SoC-independent case, which is due to that the one-segment model is not effective in reducing price volatility which in turn increases storage profits.
}
\vspace{-0.8em}

\rev{Furthermore, we also include a comparison of adopting SoC-segment models in multi-period dispatch with 1-segment model (Multi-1) and 10-segment model (Multi-10). The result shows that model SoC segments also imporve in the system cost savings as the Multi-10 is a more accurate representation of the storage model. The price results from Multi-1 and Multi-10 are similar, while Multi-1 earned more profits as we did not consider the mismatch penalty in profit calculation.}
% As shown in Fig.~\ref{fig:5a}, in SoC-dependent cases, when storage capacity is increasing, \textbf{RTD-1} first help reduces but then lifts system generation cost.  The reason is that approximating 10-segment storage nonlinear parameters to 1-segment will create more power imbalance that is costly to tackle. Simultaneously, such an approximation will create arbitrage opportunities for energy storage as shown in Fig.~\ref{fig:5e} because of over-estimating storage efficiency and charging/discharging power limits. However, as shown in Fig.~\ref{fig:5f} at high energy storage penetration rates, \textbf{RTD-1} will have a lower per-MW profit. On the other hand, although \textbf{RTD-10} produces better results than \textbf{RTD-5} and \textbf{RTD-1}, it is still not perfect as it has a large gap from the multi-period case as shown in Fig.~\ref{fig:compseg2}.}

% \todo{The averaged solving times for 24-hour real-time markets clearing are shown in Table~\ref{tab:time}. The multi case is 10-20 times faster than the RTD cases because of solving a multi-period dispatch at once, while the RTD cases are also efficient, which only consume 0.6-1.6 seconds to solve 24 single-period dispatch problems. We also notice that increasing storage capacity or segment number will not have a apparent impact on computational time.}

\begin{figure}[t]%
	\centering
    {
		\includegraphics[trim = 0mm 0mm 0mm 0mm, clip, width = .9\columnwidth]{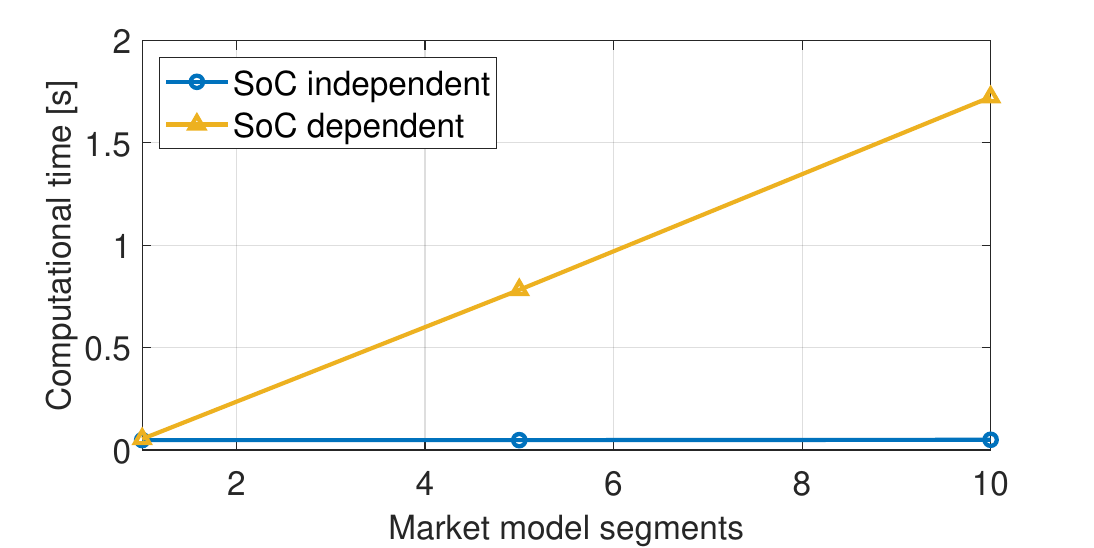}%
	}
  \caption{Solving time comparison of SoC independent and SoC dependent cases. The computational time of each segment number for 5 representative days with hourly resolution is an average of different storage capacities.}
    \label{fig:comtime}
    \vspace{-1.4em}
\end{figure}

\rev{Finally, we compare the impact of the SoC-segment model over the solution time. As shown in Fig.~\ref{fig:comtime}, in the case of SoC-independent storage models, the SoC-segment model is a linear programming model and the number of segments has negligible impacts over the computation time. However, in the case of SoC-dependent storage models, we have to introduce binary variables to enforce the segment transition logic as the underlying storage model is inherently non-convex. Thus the use of binary variables increased the solution time significantly. Overall, our result suggests the SoC-segment model improves the market efficiency and has negligible computation impact with SoC-independent storage models that are most often used in power system studies~\cite{chen2022battery}. On the other hand, while the SoC-segment market model can further improve market efficiency when extended to the more sophisticated SoC-dependent storage models, it introduces computation challenges that require careful attention when implemented in practice.}

% The averaged solving times for 24-hour real-time market clearing of RTD cases are shown in Fig.~\ref{fig:comtime}. First, as shown by the blue line in Fig.~\ref{fig:comtime}, our proposed SoC bidding model does not impact the computation time if the market clearing model run by the system operator is SoC independent, which is being used in existing markets. Second, as shown in the yellow line in Fig.~\ref{fig:comtime}, if the system operator clears the real-time markets with the multi-segment storage model, binary variables will cause significant increases in solving time. Note that this is due to the system operator intention to model inherently non-convex storage characteristics instead of our proposed solution.

% \begin{figure}[t]%
% 	\centering
% 	\subfigtopskip=2pt
% 	\subfigbottomskip=2pt
% 	\subfigcapskip=-5pt

% 	\subfigure[SoC independent and SoC dependent cases]{
% 		\includegraphics[trim = 0mm 0mm 0mm 0mm, clip, width = .9\columnwidth]{figure/Com_t_rtd2.pdf}%
% 	}
% 	\subfigure[RTD and Multi cases under SoC dependent cases]{
% 		\includegraphics[trim = 0mm 0mm 0mm 0mm, clip, width = .9\columnwidth]{figure/Com_t_multi2.pdf}
% 	}
%   \caption{Solving time comparison. The computational time of each segment number for 5 representative days with hourly resolution is an average of different storage capacities.}
%     \label{fig:comtime2}

% \end{figure}

% Such an improvement becomes increasingly apparent as storage capacity increases: At 400~MW storage capacity, the multi-segment model reduces 0.1\% of the total system operating cost, but at 1000~MW storage capacity, the multi-segment model can reduce 0.5\% of the total cost. This equals an annual cost savings of around \$2 million and \$10 million in our tested ISO-NE system, respectively. The reason is that single-segment market model approximates storage nonlinear parameters, which can generate infeasible charging/discharging commands for energy storage. As storage capacity increases, the storage power has increasing deviations from the dispatch command. Therefore, additional costs are needed to compensate such a power imbalance in frequency regulation market, resulting in higher total system operating costs. 

\section{Conclusion}\label{conc}
In this paper, we propose a new wholesale market model for energy storage that allows energy storage to submit charge and discharge bid segments according to the storage SoC ranges. Combining this model with an optimal bid generation algorithm, we show that the SoC segment market model improves storage utilization in markets from several perspectives. \rev{In the price-influencer case study, the SoC segment market model is most effective in reducing real-time prices ($\sim 10\%$) and their volatilities ($\sim 30\%$) compared to the existing storage model, as the segment model provides finer granularity for storage to respond to different prices. Using SoC segment market models can also further reduce total system cost by around 0.5\%, which is 5\% more compared to the current storage model.} The impact of the SoC segment market model on storage profit potential is mixed: When ignoring the influence of storage over market prices, the SoC segment market model provides 10\% to \rev{56\%} more profit to storage than the 1-segment model, especially if the storage parameters are sensitive to SoC; yet when considering the influence of storage on price, the SoC segment market model reduces storage profits because it is more effective at reducing system price volatilities. 

% This paper is a first effort to demonstrate how a SoC segment model could improve storage revenue potential, reduce price volatility, and improve social welfare. Many future directions remain to be explored, such as considering price prediction uncertainties, exercise of market power, and use more realistic test systems. 

The SoC segment market model provides a more accurate representation of the physical parameters and opportunity costs of energy storage. It also allows storage participants to economically manage their SoC through bid parameters. This market model opens up many new interesting research directions. \rev{The first is to incorporate the SoC segment market model into more realistic production cost models to observe the corresponding system cost, price, and storage profit estimates. Second, investigating two-stage settlement bidding strategies will be a future direction to help energy storage participants incorporate strategic bids in the day-ahead market. Besides, better reflecting SoC-dependent parameters, especially power ratings, in economic bids will be essential to narrow down the gap between SoC-independent and SoC-dependent storage models. Price uncertainty in storage bid design is another critical aspect of future research.} Most storage participants design bids based on their private price predictions; their bids will set market prices and in turn impact their future price predictions. Quantifying the relationship between market price and storage participants' price prediction strategy is crucial for analyzing future market efficiency and market power monitoring.

\bibliographystyle{IEEEtran}	% (uses file "plain.bst")
\bibliography{IEEEabrv,main}		% expects file "myrefs.bib"

% Generated by IEEEtran.bst, version: 1.14 (2015/08/26)
\begin{thebibliography}{10}
\providecommand{\url}[1]{#1}
\csname url@samestyle\endcsname
\providecommand{\newblock}{\relax}
\providecommand{\bibinfo}[2]{#2}
\providecommand{\BIBentrySTDinterwordspacing}{\spaceskip=0pt\relax}
\providecommand{\BIBentryALTinterwordstretchfactor}{4}
\providecommand{\BIBentryALTinterwordspacing}{\spaceskip=\fontdimen2\font plus
\BIBentryALTinterwordstretchfactor\fontdimen3\font minus
  \fontdimen4\font\relax}
\providecommand{\BIBforeignlanguage}[2]{{%
\expandafter\ifx\csname l@#1\endcsname\relax
\typeout{** WARNING: IEEEtran.bst: No hyphenation pattern has been}%
\typeout{** loaded for the language `#1'. Using the pattern for}%
\typeout{** the default language instead.}%
\else
\language=\csname l@#1\endcsname
\fi
#2}}
\providecommand{\BIBdecl}{\relax}
\BIBdecl

\bibitem{caiso-1}
{California ISO}, ``2020 annual report on market issues \& performance,''
  [Available Online]
  \url{http://www.caiso.com/Documents/2020-Annual-Report-on-Market-Issues-and-Performance.pdf}.

\bibitem{caiso-2}
------, ``Key statistics sep 2022 - california iso,'' [Available Online]
  \url{http://www.caiso.com/Documents/Key-Statistics-Sep-2022.pdf}.

\bibitem{caiso-3}
------, ``Public queue report - california iso,'' [Available Online, as of May
  3, 2021]
  \url{http://www.caiso.com/PublishedDocuments/PublicQueueReport.xlsx}.

\bibitem{us2021form}
{US Energy Information Association}, ``{Form EIA-860 detailed data with
  previous form data (EIA-860A/860B)},'' 2022.

\bibitem{kirschen2018fundamentals}
D.~S. Kirschen and G.~Strbac, \emph{Fundamentals of power system
  economics}.\hskip 1em plus 0.5em minus 0.4em\relax John Wiley \& Sons, 2018.

\bibitem{federal2018electric}
{Federal Energy Regulatory Commission}, ``Electric storage participation in
  markets operated by regional transmission organizations and independent
  system operators,'' \emph{Order No. 841, 162 FERC}, vol.~61, p. 127, 2018.

\bibitem{xu2020operational}
B.~Xu, M.~Korp{\aa}s, and A.~Botterud, ``Operational valuation of energy
  storage under multi-stage price uncertainties,'' in \emph{2020 59th IEEE
  Conference on Decision and Control (CDC)}.\hskip 1em plus 0.5em minus
  0.4em\relax IEEE, 2020, pp. 55--60.

\bibitem{xu2017factoring}
B.~Xu, J.~Zhao, T.~Zheng, E.~Litvinov, and D.~S. Kirschen, ``Factoring the
  cycle aging cost of batteries participating in electricity markets,''
  \emph{IEEE Transactions on Power Systems}, vol.~33, no.~2, pp. 2248--2259,
  2017.

\bibitem{pandvzic2018accurate}
H.~Pand{\v{z}}i{\'c} and V.~Bobanac, ``An accurate charging model of battery
  energy storage,'' \emph{IEEE Transactions on Power Systems}, vol.~34, no.~2,
  pp. 1416--1426, 2018.

\bibitem{fang2022efficient}
X.~Fang, H.~Guo, X.~Zhang, X.~Wang, and Q.~Chen, ``An efficient and
  incentive-compatible market design for energy storage participation,''
  \emph{Applied Energy}, vol. 311, p. 118731, 2022.

\bibitem{sakti2018review}
A.~Sakti, A.~Botterud, and F.~O’Sullivan, ``Review of wholesale markets and
  regulations for advanced energy storage services in the united states:
  Current status and path forward,'' \emph{Energy policy}, vol. 120, pp.
  569--579, 2018.

\bibitem{bhattacharjee2022energy}
S.~Bhattacharjee, R.~Sioshansi, and H.~Zareipour, ``Energy storage
  participation in wholesale markets: The impact of state-of-energy
  management,'' \emph{IEEE Open Access Journal of Power and Energy}, 2022.

\bibitem{konidena2019ferc}
R.~Konidena, ``Ferc order 841 levels the playing field for energy storage,''
  \emph{MRS Energy \& Sustainability}, vol.~6, 2019.

\bibitem{reviewall}
``Independent system operator and regional transmission organization energy
  storage market modeling working group white paper,'' [Available Online]
  \url{https://www.epri.com/research/products/000000003002012327}.

\bibitem{caiso_es}
``California iso energy storage enhancements issue paper,'' 2021, [Available
  Online]
  \url{http://www.caiso.com/InitiativeDocuments/IssuePaper-EnergyStorageEnhancements.pdf}.

\bibitem{byrne2020opportunities}
R.~H. Byrne, T.~A. Nguyen, A.~Headley, F.~Wilches-Betnal, R.~Concepcion, and
  R.~D. Trevizan, ``Opportunities and trends for energy storage plus solar in
  caiso: 2014-2018,'' in \emph{2020 IEEE Power \& Energy Society General
  Meeting (PESGM)}.\hskip 1em plus 0.5em minus 0.4em\relax IEEE, 2020, pp.
  1--5.

\bibitem{zhao2019multi}
J.~Zhao, T.~Zheng, and E.~Litvinov, ``A multi-period market design for markets
  with intertemporal constraints,'' \emph{IEEE Transactions on Power Systems},
  vol.~35, no.~4, pp. 3015--3025, 2019.

\bibitem{12akbari2019hybrid}
A.~Akbari-Dibavar, K.~Zare, and S.~Nojavan, ``A hybrid stochastic-robust
  optimization approach for energy storage arbitrage in day-ahead and real-time
  markets,'' \emph{Sustainable Cities and Society}, vol.~49, p. 101600, 2019.

\bibitem{mohsenian2015coordinated}
H.~Mohsenian-Rad, ``Coordinated price-maker operation of large energy storage
  units in nodal energy markets,'' \emph{IEEE Transactions on Power Systems},
  vol.~31, no.~1, pp. 786--797, 2015.

\bibitem{10shuai2018optimal}
H.~Shuai, J.~Fang, X.~Ai, J.~Wen, and H.~He, ``Optimal real-time operation
  strategy for microgrid: An adp-based stochastic nonlinear optimization
  approach,'' \emph{IEEE Transactions on Sustainable Energy}, vol.~10, no.~2,
  pp. 931--942, 2018.

\bibitem{13jiang2015optimal}
D.~R. Jiang and W.~B. Powell, ``Optimal hour-ahead bidding in the real-time
  electricity market with battery storage using approximate dynamic
  programming,'' \emph{INFORMS Journal on Computing}, vol.~27, no.~3, pp.
  525--543, 2015.

\bibitem{4krishnamurthy2017energy}
D.~Krishnamurthy, C.~Uckun, Z.~Zhou, P.~R. Thimmapuram, and A.~Botterud,
  ``Energy storage arbitrage under day-ahead and real-time price uncertainty,''
  \emph{IEEE Transactions on Power Systems}, vol.~33, no.~1, pp. 84--93, 2017.

\bibitem{7wang2017look}
Y.~Wang, Y.~Dvorkin, R.~Fernandez-Blanco, B.~Xu, T.~Qiu, and D.~S. Kirschen,
  ``Look-ahead bidding strategy for energy storage,'' \emph{IEEE Transactions
  on Sustainable Energy}, vol.~8, no.~3, pp. 1106--1117, 2017.

\bibitem{gao2022multiscale}
X.~Gao, B.~Knueven, J.~D. Siirola, D.~C. Miller, and A.~W. Dowling,
  ``Multiscale simulation of integrated energy system and electricity market
  interactions,'' \emph{Applied Energy}, vol. 316, p. 119017, 2022.

\bibitem{ecker2014calendar}
M.~Ecker, N.~Nieto, S.~K{\"a}bitz, J.~Schmalstieg, H.~Blanke, A.~Warnecke, and
  D.~U. Sauer, ``Calendar and cycle life study of li (nimnco) o2-based 18650
  lithium-ion batteries,'' \emph{Journal of Power Sources}, vol. 248, pp.
  839--851, 2014.

\bibitem{calero2019compressed}
I.~Calero, C.~A. Canizares, and K.~Bhattacharya, ``Compressed air energy
  storage system modeling for power system studies,'' \emph{IEEE Transactions
  on Power Systems}, vol.~34, no.~5, pp. 3359--3371, 2019.

\bibitem{saxena2019accelerated}
S.~Saxena, Y.~Xing, D.~Kwon, and M.~Pecht, ``Accelerated degradation model for
  c-rate loading of lithium-ion batteries,'' \emph{International journal of
  electrical power \& energy systems}, vol. 107, pp. 438--445, 2019.

\bibitem{jafari2019improved}
M.~Jafari, K.~Rodby, J.~L. Barton, F.~Brushett, and A.~Botterud, ``Improved
  energy arbitrage optimization with detailed flow battery characterization,''
  in \emph{2019 IEEE Power \& Energy Society General Meeting (PESGM)}.\hskip
  1em plus 0.5em minus 0.4em\relax IEEE, 2019, pp. 1--5.

\bibitem{kim2022comparison}
N.~Kim, N.~Shamim, A.~Crawford, V.~V. Viswanathan, B.~M. Sivakumar, Q.~Huang,
  D.~Reed, V.~Sprenkle, and D.~Choi, ``Comparison of li-ion battery chemistries
  under grid duty cycles,'' \emph{Journal of Power Sources}, vol. 546, p.
  231949, 2022.

\bibitem{preger2020degradation}
Y.~Preger, H.~M. Barkholtz, A.~Fresquez, D.~L. Campbell, B.~W. Juba,
  J.~Rom{\`a}n-Kustas, S.~R. Ferreira, and B.~Chalamala, ``Degradation of
  commercial lithium-ion cells as a function of chemistry and cycling
  conditions,'' \emph{Journal of The Electrochemical Society}, vol. 167,
  no.~12, p. 120532, 2020.

\bibitem{bank2022state}
T.~Bank, L.~Alsheimer, N.~L{\"o}ffler, and D.~U. Sauer, ``State of charge
  dependent degradation effects of lithium titanate oxide batteries at elevated
  temperatures: An in-situ and ex-situ analysis,'' \emph{Journal of Energy
  Storage}, vol.~51, p. 104201, 2022.

\bibitem{xu2018optimal}
B.~Xu, Y.~Shi, D.~S. Kirschen, and B.~Zhang, ``Optimal battery participation in
  frequency regulation markets,'' \emph{IEEE Transactions on Power Systems},
  vol.~33, no.~6, pp. 6715--6725, 2018.

\bibitem{koller2013defining}
M.~Koller, T.~Borsche, A.~Ulbig, and G.~Andersson, ``Defining a degradation
  cost function for optimal control of a battery energy storage system,'' in
  \emph{2013 IEEE Grenoble Conference}.\hskip 1em plus 0.5em minus 0.4em\relax
  IEEE, 2013, pp. 1--6.

\bibitem{yang2022multi}
W.~Yang, Y.~Wen, H.~Pand{\v{z}}i{\'c}, and W.~Zhang, ``A multi-state control
  strategy for battery energy storage based on the state-of-charge and
  frequency disturbance conditions,'' \emph{International Journal of Electrical
  Power \& Energy Systems}, vol. 135, p. 107600, 2022.

\bibitem{zheng2015study}
Y.~Zheng, M.~Ouyang, L.~Lu, J.~Li, Z.~Zhang, and X.~Li, ``Study on the
  correlation between state of charge and coulombic efficiency for commercial
  lithium ion batteries,'' \emph{Journal of power Sources}, vol. 289, pp.
  81--90, 2015.

\bibitem{caiso2020esproposal}
\BIBentryALTinterwordspacing
{California ISO}, ``Energy storage and distributed energy resources phase 4
  proposal - final proposal.'' [Online]. Available:
  \url{http://www.caiso.com/InitiativeDocuments/FinalProposal-EnergyStorage-DistributedEnergyResourcesPhase4.pdf}
\BIBentrySTDinterwordspacing

\bibitem{chen2022convexifying}
C.~Chen and L.~Tong, ``Convexifying market clearing of soc-dependent bids from
  merchant storage participants,'' \emph{arXiv preprint arXiv:2209.02107},
  2022.

\bibitem{zheng2022comparing}
N.~Zheng and B.~Xu, ``Impact of bidding and dispatch models over energy storage
  utilization in bulk power systems,'' \emph{IREP Symposium on Bulk Power
  System Dynamics and Control 2022}, 2022.

\bibitem{castillo2013profit}
A.~Castillo and D.~F. Gayme, ``Profit maximizing storage allocation in power
  grids,'' in \emph{52nd IEEE Conference on Decision and Control}.\hskip 1em
  plus 0.5em minus 0.4em\relax IEEE, 2013, pp. 429--435.

\bibitem{krishnamurthy20158}
D.~Krishnamurthy, W.~Li, and L.~Tesfatsion, ``An 8-zone test system based on
  iso new england data: Development and application,'' \emph{IEEE Transactions
  on Power Systems}, vol.~31, no.~1, pp. 234--246, 2015.

\bibitem{chen2022battery}
Y.~Chen and R.~Baldick, ``Battery storage formulation and impact on day ahead
  security constrained unit commitment,'' \emph{IEEE Transactions on Power
  Systems}, 2022.

\end{thebibliography}

\begin{appendices}

\section*{Appendix}

\subsection{Unit Commitment Formulation}\label{app:uc}
\begin{subequations}
The objective function of unit commitment minimizes daily generation costs of thermal generators:
\begin{align}
    \min \sum_{t=1}^{T}\sum_{i=1}^{N_g} C\up{l}_i g_{i,t} + C\up{q}_i g^2_{i,t} + C\up{n}_{i}u_{i,t} + C\up{s}_{i} v_{i,t}
\end{align}

Decision variables include continuous variables $g_{i,t}$, $ r_{i,t}$, and $ w_t $; binary variables $u_{i,t}$, $v_{i,t}$, $z_{i,t}$.
$g_{i,t}$ are the electric power generation of thermal generator $i$ at period $t$, $u_{i,t}$ is a binary variable indicating whether generator $i$ is on at period $t$, and $z_{i,t}$ is a binary variable indicating whether generator $i$ turns off at period $t$. $C\up{l}_i$ and $C\up{q}_i$ are the first and second order terms for the marginal production cost of generator $i$, $C\up{n}_{i}$ is the no load cost, and $C\up{s}_{i}$ is the startup cost.
$N_g$ is the number of thermal generators and $T$ is the number of steps. 

The power generation of thermal generators should satisfy generation limits: 
\begin{align}
    \mathrm{Gmin}_i \cdot u_{i,t} \leq g_{i,t} \leq \mathrm{Gmax}_i \cdot u_{i,t} \label{glimit} \ 
\end{align}
where $\mathrm{Gmin}_i$ and $\mathrm{Gmax}_i$ denotes the minimum and maximum generation of thermal generator $i$. 

% Moreover, thermal generator must also satisfy ramping constraints: The increment or decrement of power generations between two consecutive time periods should not exceed their ramping rates:
% \begin{align}
%     -\mathrm{RR}_i\leq g_{i,t}-g_{i,t-1} \leq \mathrm{RR}_i + \mathrm{Gmin}_i \cdot y_{i,t}, \quad t\in\{1,2,\dotsc, T\}  \label{gramp}\
% \end{align}
% where  $\mathrm{RR}_i$ is ramp rate of generator $i$. Scalar $y_{i,t}$ is a binary variable indicating whether generator $i$ turns on at period $t$. 

Startup and shutdown logic constraints:
\begin{align}
    y_{i,t} - z_{i,t} &= u_{i,t} - u_{i,t-1} \\
    y_{i,t} + z_{i,t} &\leq 1 
\end{align}

Generator minimum up and down time constraints:
\begin{align}
    \textstyle \sum_{\tau = \max\{t - \mathrm{Tup}_i + 1, 1\}}^t y_{i,\tau} &\leq u_{i,t}  \\
    \textstyle \sum_{\tau = \max\{t - \mathrm{Tdn}_i + 1,1\}}^t z_{i,\tau} &\leq 1-u_{i,t}
\end{align}
where $\mathrm{Tup}_i$ and $\mathrm{Tdn}_i$ are maximum up time and minimum down time of generator $i$, respectively.
 
Reserve constraints following the 5+3 rule (5\% renewables and 3\% demand):
\begin{align}
    \textstyle \sum_{i=1}^{N_g} r_{i,t} &\geq (5\%)w_t + (3\%) D_t \\
    r_{i,t} &\leq \mathrm{Gmax}_i u_{i,t} - g_{i,t} 
    % r_{i,t} &\leq RR_i
\end{align}
where $w_{t}$ is accommodated wind generation during time period $t$.
The wind generation should satisfy:
\begin{align}
    w_t \leq \tilde{W}_{t} \label{wind}\
2\end{align}
where $\tilde{W}_{t}$ denotes the day-ahead wind generation forecast at period $t$.

The electric power balance constraint and price:
\begin{align}
    \textstyle \sum_{i=1}^{N_g} g_{i,t} + w_t  = D_t : \lambda\up{DA}_{t} 
\end{align}\label{gbalance}\
where the dual variable associated with constraint~\eqref{gbalance} is the day-ahead price $\lambda_{DA,t}$ at period $t$. 
\end{subequations}

\subsection{Economic Dispatch}\label{app:ed}
The \rev{multi-period} economic dispatch problem has the following objective function:
\begin{subequations}
\begin{align}
    \min \sum_{t=1}^{T}\sum_{i=1}^{N_g} C\up{l}_i g_{i,t} + C\up{q}_i g^2_{i,t} + \sum_{t=1}^{T}\sum_{s=1}^S c_{s} d_{s,t}
    \label{gbalance1}
\end{align}
subject to the power balance constraint:
\begin{align}
    \textstyle \sum_{i=1}^{N_g} g_{i,t} + w_t + \sum_{s=1}^S d_{s,t} = D_t + \sum_{s=1}^S p_{s,t} : \lambda\up{RT}_{t} \label{gbalance2}
\end{align}
which now includes storage charge and discharge power from all segments. Other constraints include generator ratings \eqref{glimit}, wind limits \eqref{wind}, and storage unit constraints \eqref{p1_c1}--\eqref{p1_c3}.

The single period economic dispatch problem is:
\begin{gather}
\min \sum_{i=1}^{N_g} C\up{l}_i g_{i,t} + C\up{q}_i g^2_{i,t} + \sum_s^S( G_{t,s}d_{t,s} - B_{t,s}p_{t,s}) \label{gbalance3}
\end{gather}
subject to same constraints as the multi-period dispatch problem \eqref{gbalance2}, \eqref{glimit},  \eqref{wind},  \eqref{p1_c1}--\eqref{p1_c3}. Decision variables in both problems include  $g_{i,t}$, $ w_t $, $p_{t,s}$, $d_{t,s}$. Note that the commitment status $u_{i,t}$ is from the unit commitment results.

\end{subequations}
\end{appendices}
\end{document}